\documentclass[preprintnumbers, floatfix, letterpaper, aps, prd, epsfig, nofootinbib,
twocolumn
]{revtex4-1}
\usepackage{bm, graphicx, dcolumn, epstopdf, epsf, latexsym, mathbbol, amssymb, amsmath, color, slashed, mathrsfs, mathcomp, simplewick}
\usepackage[center]{subfigure}
\usepackage{multirow}
\usepackage{makecell}
\usepackage[colorlinks,linkcolor=blue,citecolor=blue,urlcolor=blue]{hyperref}

\allowdisplaybreaks
\begin{document}

\title{\textbf{Matter accretion onto the magnetically charged Euler-Heisenberg black hole with scalar hair }}
\author{ H. Rehman $^1$
\footnote{hamzarehman244@gmail.com}, G. Abbas $^{1,2}$ \footnote{ghulamabbas@iub.edu.pk}, Tao Zhu $^{3, 4}$ \footnote{zhut05@zjut.edu.cn}, and G. Mustafa $^{5,6}$ \footnote{gmustafa3828@gmail.com} }
\address{ ${}^1$ Department of Mathematics, The Islamia University of Bahawalpur, Bahawalpur Pakistan.\\
${}^2$ National Astronomical Observatories, Chinese Academy of Sciences, Beijing 100101, China\\
${}^3$ Institute for Theoretical Physics \& Cosmology, Zhejiang University of Technology, Hangzhou, 310023, China\\
${}^{4}$ United Center for Gravitational Wave Physics (UCGWP),  Zhejiang University of Technology, Hangzhou, 310023, China\\
${}^5$ Department of Physics, Zhejiang Normal University, Jinhua 321004, People's Republic of China\\
${}^6$New Uzbekistan University, Mustaqillik Ave. 54, Tashkent 100007, Uzbekistan}

\date{\today}

\begin{abstract}

This paper deals with astrophysical accretion onto the magnetically charged Euler-Heisenberg black holes with scalar hair. We examine the accretion process of a variety of perfect fluids, including polytropic and isothermal fluids of the ultra-stiff, ultra-relativistic, and sub-relativistic forms, when fluid is accreting in the vicinity of the black hole. By using the Hamiltonian dynamical approach, we can find the sonic or critical points numerically for the various types of fluids that are accreting onto the black hole. Furthermore, for several types of fluids, the solution is provided in closed form, expressing phase diagram curves. We compute the mass accretion rate of a magnetically charged Euler-Heisenberg black hole with scalar hair. We observe that the maximum accretion rate is attained for small values of the black hole parameters. We may be able to understand the physical mechanism of accretion onto black holes using the outcomes of this investigation.


\end{abstract}
\maketitle
\section{Introduction}

We commence by emphasizing that one of the most unexpected events in our universe is the existence of black holes (BHs). We want to have a discussion about BH as an aspect of classical physical theory. Einstein's theory of gravity, which establishes time as well as space, is the classical theory. Black holes were first thought to exist exclusively in theory, and while their models were researched in considerable detail, many scientists, including Einstein, questioned whether or not they actually existed. How does Einstein's theory of gravity explain the concept of time and space around a huge object like a star? is a natural question to ask in this situation. Schwarzschild discovered the answer to this problem, which simply relies on the mass of all static round objects. Yet, when all of the mass is contained inside a specific radius known as the Schwarzschild radius, surprising things may happen. Then, as an event horizon occurs at the Schwarzschild radius, the term "BH" is used.

It is generally accepted that celestial objects, such as BHs, gather mass through a process known as accretion. The accretion phenomenon that surrounds enormous gravitating objects is a fundamental concept in astrophysics that plays a vital role in comprehending a number of astrophysical behaviors and speculations, such as the formation of super-massive (BHs), the expansion of stars, the emission of X-rays from compact star binaries, the luminosity of quasars, and other phenomena \cite{b7}-\cite{b9}. The accretion of matter in a realistic astrophysical procedure is incredibly complex since it encompasses numerous significant problems of general relativistic magnetohydrodynamics, such as nuclear burning, turbulence, radiation processes, etc. It is useful to summarize the challenge by establishing some assumptions or assuming some basic conditions in order to comprehend the general accretion processes.

The Bondi stationary, spherically symmetric solution \cite{c1} illustrates the fundamental accretion process by depicting an infinitely massive homogenous gas cloud steadily accreting onto a gravitational object in the center. Newtonian gravity serves as the foundation for the Bondi approach. Thereafter, Michel \cite{c2} examined the steady-state spherically symmetric flow of test fluids towards a Schwarzschild BH within the context of general relativity (GR). Subsequently, Shapiro and Teukolsky \cite{c3} also made contributions to the concept of relativistic accretion on compact objects. Furthermore, Babichev et al. \cite{c4} found that if phantom energy is permitted to accrete onto the BH throughout the accretion procedure, the BH mass may drop. Additionally, as demonstrated by Jamil et al. \cite{c5} the phantom accretion not only reduces the BH mass but also transforms it into a naked singularity. Debnath \cite{b6} elaborated the static accretion onto a general class of spherically symmetric BHs by examining the impact of the cosmological constant on the accretion rate, in accordance with the Babichev model. Bondi-type accretion onto the Reissner-Nordstrom anti-de-sitter spacetime was studied by Ficek \cite{c7}. Using the methodology described in Ficek \cite{c7}, Ahmed et al. \cite{a8} investigated the process of accretion onto the Reisner-Nordstrom anti-de-sitter BH with a global monopole. In the $f (R)$ and $f (T)$ modified theories of gravity, they elaborated their prior research for accretion onto BH \cite{a9, b1}. Abdul Jawad and M. Umair Shahzad have investigated the accreting fluids onto regular BH using the Hamiltonian Technique \cite{b5}. Matter Accretion onto a Conformal Gravity BH was determined by G. Abbas and A. Ditta \cite{b51}. In \cite{b52}, Astrophysical accretion near a regular Hayward BH was calculated by A. Ditta and G. Abbas. The accretion of matter onto a brane-world BH through the Hamiltonian approach has been described in Ref. \cite{b53}. The research conducted by Sen Yang et al. focuses on the examination of the spherical accretion flow onto general parameterized spherically symmetric BH space-times \cite{b54}. Moreover, the spherical accretion is discussed in \cite{b55,b56}. General relativistic dust accretion for stationary rotating BHs was established by Azreg-Ainou \cite{c8}. In literature \cite{d1}- \cite{d6} have discussed the accretion phenomena in numerous space-times.

The process of transonic accretion and the presence of the sonic point (or critical point) are significant aspects of spherical accretion onto the BH. During the accretion flow transitions occurs from a subsonic to a supersonic state at the sonic point. In a particular BH space-time, the sonic points are typically found close to the horizon. The narrow region near the sonic point is significant and fascinating, it is necessarily linked to current research on the gravitational and electromagnetic wave spectra. Consequently, investigation of the spherical accretion problem can not only help in our understanding of the accretion process in various BH but also, more significantly, convey a distinct perspective on how to investigate the nature of BH space-time under strong gravity.

This paper mainly focuses on the analysis of astrophysical accretion near the magnetically charged Euler-Heisenberg (EH) BH with scalar hair.
The electrodynamics EH Lagrangian was first proposed in 1936 \cite{d7}. In \cite{d9}, a methodology for identifying the impact of the EH hypothesis was presented. It was only the natural way to relate the EH Lagrangian to the Ricci scalar via the volume element to investigate BH solutions because the EH theory possesses remarkable physical characteristics. In \cite{e1}, analytical solutions were found for the magnetically charged situation while simultaneously discussing electric charges and dyons. This work provided the first BH solution to EH electrodynamics. In \cite{e2} and \cite{e3}, electrically charged BH was addressed, whereas the geodesic structure was the focus of the research in \cite{e3}. In \cite{e4} the authors have examined the charged particle motions surrounding the EH AdS BH. Whereas the quasi-normal modes were computed in \cite{e5}, the thermodynamics of these BH were explored in \cite{e6, e7}. The BH together with the EH Lagrangian and modified gravity theories were investigated in \cite{f1}-\cite{f3}. Ultimately, the shadow of EH BH was studied in \cite{f4}. Recently, a BH solution is obtained in the EH theory \cite{f5}. The effects of the coupling constant of EH theory on the thermodynamics and energy conditions of this BH have also been investigated \cite{f5}. Also, the motion of particles in a magnetically charged EH BH with scalar hair is studied in \cite{f6}. The fundamental objective of the present investigation is to address the obvious question of whether magnetically charged EH BH with scalar hair might affect astrophysical accretion processes using the Hamiltonian approach. Focusing on perfect fluid accretion onto the magnetically charged EH BH space-times, we investigate the transonic phenomena for various categories of fluid, including isothermal fluids (such as ultra-stiff, ultra-relativistic, radiation, and sub-relativistic) and polytropic fluid.

The following is the structure of our paper: A brief introduction to the magnetically charged EH BH with scalar hair is given in Sec. II. In Sec. III, we provide some helpful quantities and construct the fundamental equations for subsequent consideration of the spherical accretion of various fluids. We examine the accretion processes and determine the critical points of the system in Sec. IV. We apply the obtained formalism or findings to a number of well-known fluids and extensively investigate the transonic phenomenon for the accretion of these fluids in the magnetically charged EH BH with scalar hair in sec. V. Furthermore, we compute the BH mass accretion rate in Sec. VI by considering the impact of the accelerating parameters. Finally, we provide a summary of this article in Sec. VII.

\section{The Space-time of Magnetically Charged EH Black Holes with Scalar Hair}

This section provides a concise overview of magnetically charged EH BH in the Einstein-Euler-Heisenberg theory. For this purpose, the Euler-Heisenberg action in the presence of a scalar field is given by \cite{f5},
\begin{eqnarray}
S&=&\int d^{4}x\sqrt{-g}\mathcal{L} \nonumber\\
&=&\int d^{4}x\sqrt{-g}\Big(\frac{R}{2}-\frac{1}{2}\partial ^{\mu}\phi \partial _{\mu}\phi-V(\phi) \nonumber\\
&&~~~~~~~~~~~~~~~~~~  -P+ \alpha P^{2}+\beta Q^{2} \Big),
\end{eqnarray}
 where $\mathcal{L}$ represents the Lagrangian of the Einstein-Euler-Heisenberg theory, $R$ is the Ricci scalar, $P=F_{\mu \nu} F^{\mu \nu}$, $Q=\epsilon_{\mu \nu \rho \sigma}F_{\mu \nu}F^{\rho \sigma}$, $F_{\mu \nu}=\partial _{\mu}A_{\nu}-\partial _{\nu}A_{\mu}$ is the field strength, and $\epsilon_{\mu \nu \rho \sigma}$ denotes the Levi-Civita tensor that fulfill
\begin{equation}
\epsilon_{\mu \nu \rho \sigma} \epsilon^{\mu \nu \rho \sigma}=-24.
\end{equation}
The corresponding field equations can be obtained by varying the above action with respect to the spacetime metric $g_{\mu\nu}$, the scalar field $\phi$, and the electromagnetic field $A_\mu$, which are given respectively by
\begin{eqnarray}
&&G_{\mu \nu}= T_{\mu \nu}\equiv T^{\phi}_{\mu \nu}+T^{EM}_{\mu \nu},\label{j1} \\
&&\Box\phi= \frac{dV}{d\phi},\label{j2}\\
&&\nabla_{\mu}(F^{\mu \nu}-2 \alpha PF^{\mu \nu}-2\beta Q\epsilon^{\mu \nu \rho \sigma} F_{\rho \sigma})=0,\label{j3}
\end{eqnarray}
where
\begin{eqnarray}
&&T^{\phi}_{\mu \nu}=\partial_{\mu}\phi\partial_{\nu}\phi-\frac{1}{2}g_{\mu\nu}\partial^{\alpha}\phi\partial_{\alpha}\phi-g_{\mu\nu}V(\phi),\label{j4}\\
&&T^{Em}_{\mu \nu}=2F_{\mu \rho}F^{\rho}_{\nu}+\frac{1}{2}g_{\mu\nu}(-P+\alpha P^{2}+\beta Q^{2}) \nonumber\\
&&~~~~~~~~ -4\alpha PF_{\mu \rho}F^{\rho}_{\nu}-8\beta Q\epsilon_{\mu\xi\eta\rho} F^{\xi\eta}F^{\rho}_{\nu}.\label{j5}
\end{eqnarray}

The mentioned spherically symmetric space-time metric ansatz is taken into consideration as follows
\begin{equation}
ds^{2}= -b(r)dt^{2}+ \frac{1}{b(r)}dr^{2}+ b_{1}(r)^{2}(dr^{2}+\sin^{2}\theta d\phi^{2}),\label{a1}
\end{equation}
by assuming the four-vector $A_{\mu}$
\begin{equation}
A_{\mu}= \big(\mathcal{A}(r), 0, 0, {Q}_{m} \cos\theta \big),
\end{equation}
where ${Q}_{m}$ represents the magnetic charge for the BH. Also, we define the following quantities
\begin{eqnarray}
P=\frac{2{Q}_{m}^{2}}{b_{1}(r)^{4}}-2\mathcal{A}'(r)^{2},\\
Q=-\frac{8 {Q}_{m}\mathcal{A}'(r)}{b_{1}(r)^{2}}.
\end{eqnarray}
If dyons are not taken into account (both magnetic and electric charge), $Q$ will become extinct. By using Eqs. (\ref{j1}) to (\ref{j5}), we obtain
\begin{eqnarray}
\mathcal{A}(r)=0,\\
\phi(r)=\frac{1}{\sqrt{2}}\ln\left(1+\frac{\nu}{r}\right),
\end{eqnarray}
and the metric function $b_1(r)$ and $ b(r)$ are given by
\begin{eqnarray}
b_{1}(r)&=&\sqrt{r(\nu +r)}, \\
b(r)&=& c_{1}r(r+\nu)+\frac{(2r-c_{2})(\nu+2r)-4{Q}_{m}^{2}}{\nu^{2}} \nonumber\\
&&+\frac{8\alpha {Q}_{m}^{4}(12r^{2}+12r\nu-\nu^{2})(\nu^{2}+3r^{2}+3r\nu)}{3r^{2}\nu^{6}(r+\nu)^{2}}\nonumber\\
&&\nonumber+\frac{2}{\nu^{8}}\ln\left(\frac{r}{r+\nu}\right)\Bigg[- \nu^{5}r(r+\nu)(\nu+c_{2}) \nonumber\\
&&-2{Q}_{m}^{2}(r+\nu)(\nu^{4}-24\alpha {Q}_{m}^{2})\ln\left(\frac{r}{r+\nu}\right)\nonumber\\
&&+48\nu\alpha {Q}_{m}^{4}{(2r+\nu)}-2{Q}_{m}^{2}\nu^{5}{(2r+\nu)}\Bigg],
\end{eqnarray}
where $ c_{1}$ and $ c_{2}$ represent constants of integration and $\nu$ is used to analyze the behavior of the scalar charge. At sufficiently large distances, we will impose that $\nu> 0$. Hence, we have
\begin{eqnarray}
b(r\rightarrow\infty)& \sim & 1 -\frac{-\nu-c_{2}}{3r}+\frac{\nu^2+\nu c_{2}+6{Q}_{m}^{2}}{6r^2} \nonumber\\
&&+r^2\left(c_{1}+\frac{4}{\nu^2}\right)+\frac{r(4+\nu^2 c_{1})}{\nu}\\
&&-\frac{\nu(\nu c_{2}+10{Q}_{m}^{2}+\nu^2)}{10 r^{3}}+O\left(\frac{1}{r^{4}}\right). \label{b_inf}
\end{eqnarray}
A new scale is introduced into the theory by the scalar charge, yielding an underlying cosmological constant. Since the generated mass term is determined by the scalar charge and an integration constant, the BH has a secondary scalar hair, so
\begin{eqnarray}
b(r\rightarrow\infty)& \sim &  1 -\frac{2m}{r}+\frac{m\nu+{Q}_{m}^{2}}{r^2}-\frac{r^2\Lambda _{\text{eff}}}{3} \nonumber\\
&& -\frac{r}{3}\Lambda _{\text{eff}} \nu-\frac{\nu (3m\nu+5{Q}_{m}^{2})}{5 r^{3}}+ O\left(\frac{1}{r^{4}}\right). \nonumber\\
\end{eqnarray}
where $m=\frac{\nu +c_{2}}{6}$ and $ \Lambda_{\text{eff}}=-(3c_{1}+\frac{12}{\nu^2})$.
The condition for the small scalar case is achieved when $\nu\rightarrow 0$, thus
\begin{eqnarray}
b&=&\left(1-\frac{2 m}{r}+\frac{Q_m^2}{r^2}-\frac{2 \alpha Q_m^4}{5 r^6}-\frac{\Lambda _{\text{eff}}r^2}{3}\right) \nonumber\\
&& +\nu \left(\frac{m}{r^2}+\frac{6 \alpha Q_m^4}{5 r^7}-\frac{Q_m^2}{r^3}-\frac{\Lambda _{\text{eff}}r}{3}\right). \label{a01}
\end{eqnarray}
Here $m$, ${Q}_{m}$, $\alpha$, $\Lambda _{\text{eff}}$ and $\nu$ represents the mass, magnetic charge, EH parameter, the cosmological constant, and the coupling small scalar hair charged parameter of the EH BH, respectively.
By setting $b = 0$, one can determine the horizon of the BH. By using Eq.~(\ref{b_inf}), we are unable to determine the horizon of BH analytically, so it is expressed graphically in \textbf{Fig. 1}.

\begin{figure}
\includegraphics[width=8.0cm]{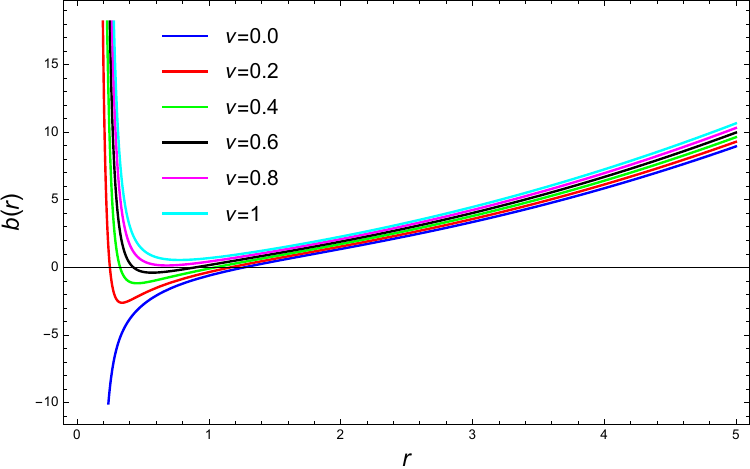}
\caption{The graph of horizon radius for $m=1$, $Q_{m}=0.2$, $\Lambda _{\text{eff}}=-1$, $\alpha=1$ and various values of $\nu$}.
\end{figure}

\section{ Fundamental equations for spherical accretion flow}

In this section, we calculate the fundamental equations of accretion around the magnetically charged EH BH with small scalar hair. To analyze this, we use two fundamental laws: the conservation of the number of particles and the conservation of energy. We assume perfect fluid is flowing around the BH. The energy-momentum tensor for the perfect fluid is given by
\begin{equation}
T^{\mu\nu}=(e+p)u^{\mu}u^{\nu}-pg^{\mu\nu},\label{a2}
\end{equation}
where $e$ and $p$ represent the energy density and pressure, respectively. If the proper number density is $n$, then the flux density is defined by $J^{\mu}=nu^{\mu}$, where $u^{\mu}=\frac{dx^{\mu}}{d\tau}$ is the 4-velocity of the particles. In the accretion process, we presume that no particles are formed or destroyed, which means that the total number of particles is conserved, so particle conservation and energy conservation are given as follows
\begin{equation}
\nabla_{\mu}J^{\mu}= \nabla_{\mu}(nu^{\mu})=0,\label{a3}
\end{equation}
\begin{equation}
\nabla_{\mu}T^{\mu\nu}=0.\label{a4}
\end{equation}
By solving Eq. (\ref{a3}) in equatorial plane $(\theta=\frac{\pi}{2})$, we have
\begin{equation}
r (\nu +r)nu=C_{3},\label{a5}
\end{equation}
where $C_{3}$ is an integration constant. Since we consider the flow of the fluid to be in a radial direction in the equatorial plane, only two components $u^{t}$ and $u^{r}=u$  are different from zero. By utilizing the normalization condition, we attain
\begin{equation}
(u^{t})^{2}=\frac{{\cal F}(r)+u^{2}}{{\cal F}^2(r)} \label{a6}
\end{equation}
with ${\cal F}$ being given by
\begin{eqnarray}
{\cal F}(r)&\equiv& 1-\frac{2 m}{r}+\frac{Q_m^2}{r^2}-\frac{2 \alpha Q_m^4}{5 r^6}-\frac{\Lambda _{\text{eff}}r^2}{3} \nonumber \\
&&+\nu\left(\frac{m}{r^2}+\frac{6 \alpha Q_m^4}{5 r^7}-\frac{Q_m^2}{r^3}-\frac{\Lambda _{\text{eff}}r}{3}\right),
\end{eqnarray}
and $u_{t}$ takes the form
\begin{equation}
u_{t}=\sqrt{{\cal F}(r)+u^{2}}.\label{a7}
\end{equation}
Moreover, the first law of thermodynamics for a perfect fluid is stated as \cite{a21}
\begin{equation}
dp=n(dh-Tds), \,\,\,\,\,\,\,\ de=hdn+nTds, \label{a8}
\end{equation}
where $s$ represents entropy, $T$ indicates the temperature, and $h$ denotes the specific enthalpy which is denied by
\begin{equation}
h=\frac{e+p}{n}. \label{a9}
\end{equation}
There is a scalar $hu_{\mu}\xi^{\mu}$ in relativistic hydrodynamics that is conserved along the fluid's trajectories \cite{a21}, so
\begin{equation}
u^{\nu}\nabla_{\nu} (hu_{\mu}\xi^{\mu})=0, \label{a9}
\end{equation}
where $\xi^{\mu}$ stands for the Killing vector of space-time. If we assume $\xi^{\mu}=(1,0,0,0)$, we attain
\begin{equation}
\partial_{r}(hu_{t})=0. \label{b1}
\end{equation}
Integrating the above equation, we have
\begin{equation}
h  \sqrt{{\cal F}(r)+u^{2}}=C_{4}, \label{b2}
\end{equation}
where $C_{4}$ is integration constant. It is simple to demonstrate that the specific entropy of the fluid is conserved across the flow lines $u^{\mu}\nabla_{\mu}s=0$. If we rewrite $T^{\mu\nu}$ as $nhu^{\mu}u^{\nu}+(nh-e)g^{\mu\nu}$ and then apply the conservation equation of $T^{\mu\nu}$ onto $u^{\mu}$, we get
\begin{eqnarray}\nonumber
u_{\mu}\nabla_{\mu}T^{\mu\nu}&=&u_{\mu}\nabla_{\mu}(nhu^{\mu}u^{\nu}+(nh-e)g^{\mu\nu})
\\ &=&u^{\mu}(h\nabla_{\mu}n-\nabla_{\mu}e)\nonumber \\
&=&-nTu^{\mu}\nabla_{\mu}s=0. \label{b3}
\end{eqnarray}
In the specific scenario, we assume that the motion of the fluid is radial, static (it does not change over time), and it conserves the BH spherical symmetry, so the above equation reduces to $\partial_{r} s = 0$, which means that $s$ is constant everywhere. As a result, the fluid's motion is isentropic and
 Eq. (\ref{a8}) becomes
\begin{equation}
dp=ndh, \,\,\,\,\,\,\,\ de=hdn, \label{b4}
\end{equation}
we will study the flow by using Eqs. (\ref{a5}), (\ref{b2}) and (\ref{b4}). Also, it transforms the equation of state (EOS) of a simple fluid, $e = e (n, s)$, from its canonical form to its barotropic form because $s$ is a constant, so
\begin{equation}
e=F(n), \label{b5}
\end{equation}
by using second Eq. (\ref{b4}), we obtain $h=\frac{de}{dn}$, which gives us
\begin{equation}
h=F'(n), \label{b6}
\end{equation}
here, $'$ indicates the derivative with respect to $n$. Also, $p' = nh'$ is produced by the first equation of Eq. (\ref{b4}), when $h = F'(n)$, we get
\begin{equation}
p'=nF''(n), \label{b7}
\end{equation}
by integrating Eq. (\ref{b7}), we have
\begin{equation}
p=nF'(n)-F(n). \label{b8}
\end{equation}
We know that the EOS of the form $p = G (n)$ cannot exist independently of the EOS of the form $e=F (n)$. By solving the above differential equation, the connection between $F$ and $G$ can be determined.
\begin{equation}
G(n)=nF'(n)-F(n). \label{b9}
\end{equation}
We can determine the sound speed in a local inertial frame by using the following formula $a^{2}=(\frac{\partial p}{\partial e})_{s}$,
which is given in \cite{a5}. Because entropy $s$ is constant, we can reduce this to $a^{2} = dp/de$. From Eq. (\ref{b4}), we determine the useful formula for the subsequent parts as follows
\begin{equation}
a^{2}=\frac{dp}{de}=\frac{ndh}{hdn}\Rightarrow \frac{dh}{h}=a^{2}\frac{dn}{n}, \label{c1}
\end{equation}
utilizing Eq. (\ref{b4}), in Eq. (\ref{c1}), we have
\begin{equation}
a^{2}=\frac{ndh}{hdn}= \frac{n}{F'}F''=n (\ln F')'.\label{c2}
\end{equation}
Another useful expression is the three-dimensional fluid velocity $\omega$, determined by a local stationary observer.
As the motion in the equatorial plane is radial, so $d\theta = d\phi= 0$, and then Eq. (\ref{a1}) becomes
\begin{eqnarray}
ds^2= - {\cal F}(r) dt^2 + \frac{dr^2}{{\cal F}(r)}.
\end{eqnarray}
In the usual relativistic method \cite{a3, a4} as seen by a local, stationary observer, the typical three-dimensional velocity $\omega$ can be defined as
\begin{eqnarray}
\omega= \frac{1}{{\cal F}(r)} \frac{dr}{dt},
\end{eqnarray}
which yields
\begin{eqnarray}
\omega^2 = \frac{u^2}{{\cal F}(r)+u^2}.
\end{eqnarray}
By using
$u_{t}=- {\cal F}(r)u^{t}$, $u^{r}=u=\frac{dr}{d\tau}$, $u^{t}=\frac{dt}{d\tau}$, and Eq. (\ref{a7}), we have
\begin{equation}
u^{2}=\frac{\omega^{2}}{1-\omega^{2}} {\cal F}(r), \label{c6}
\end{equation}
and
\begin{equation}
u_{t}^{2}=\frac{{\cal F}(r)}{1-\omega^{2}}. \label{c8}
\end{equation}
Then employing Eq. (\ref{a5}), we have
\begin{widetext}
\begin{equation}
\frac{n^{2}\omega^{2}(\nu +r)^2 \Big[5 r^5 \left(m (6 r-3 \nu )+r^2 (\Lambda_{\rm eff}  r (\nu +r)-3)\right)+6 \alpha  Q_m^4 (r-3 \nu )+15 Q_m^2 r^4 (\nu -r)\Big]}{15r^{5}(1-\omega^{2})}=C_{3}^{2}. \label{c9}
\end{equation}
\end{widetext}
These equations will be utilized in the following Hamiltonian evaluation.

\section{Dynamical system and sonic points}

According to the fundamental Eqs. (\ref{a5}) and (\ref{b2}), there are integration constants $C_{3}$ and $C_{4}$. Also, Hamiltonian $\mathcal{H}$ is read as the square of the L.H.S side of Eq. (\ref{b2}), which is given by
\begin{eqnarray}
\mathcal{H}&=&h^{2}\Bigg[1-\frac{2 m}{r}+\frac{Q_m^2}{r^2}-\frac{2 \alpha Q_m^4}{5 r^6}-\frac{\Lambda _{\text{eff}}r^2}{3}\nonumber \\
&&+\nu\left(\frac{m}{r^2}+\frac{6 \alpha Q_m^4}{5 r^7}-\frac{Q_m^2}{r^3}-\frac{\Lambda _{\text{eff}}r}{3}\right)+u^{2}\Bigg].\label{d1}
\end{eqnarray}
Here, we establish the Hamiltonian dynamical system as a function of $(r, \omega)$ to study the Michal flow, which is found in Chaverra et al. \cite{a6, a7}, Ahmed et al. \cite{a8}-\cite{b1}, and is written in the given scenario as
\begin{eqnarray}
\mathcal{H}(r,\omega) &=&\frac{h^{2}(r,\omega)}{1-\omega^{2}} \Bigg[1-\frac{2 m}{r}+\frac{Q_m^2}{r^2}-\frac{2 \alpha Q_m^4}{5 r^6}-\frac{\Lambda _{\text{eff}}r^2}{3}\nonumber \\
&& +\nu\left(\frac{m}{r^2}+\frac{6 \alpha Q_m^4}{5 r^7}-\frac{Q_m^2}{r^3}-\frac{\Lambda _{\text{eff}}r}{3}\right)\Bigg].\label{d2}
\end{eqnarray}
Furthermore, the dynamical system corresponding to the Hamiltonian is expressed as follows
\begin{equation}
\dot{r}=\mathcal{H}_{,\omega}, \,\,\,\,\, \dot{\nu}=\mathcal{H}_{,r} \label{d3}
\end{equation}
where the $\bar{t}$-derivative is represented by dot. According to Eq. (\ref{d3}), $\mathcal{H}_{,\omega}$ is the partial derivative of $\mathcal{H}$ with respect to $\omega$ when $r$ is assumed to be constant, and $\mathcal{H}_{,r}$ stands for the partial derivative of $\mathcal{H}$ with respect to $r$ when $\omega$ is taken to be constant. Ultimately, the system (\ref{d3}), reduces to
\begin{eqnarray}
\dot{r}&=&\frac{2\omega (\omega^{2}-a^{2})h^{2}}{(1-\omega^{2})^{2}}\Bigg[1-\frac{2 m}{r}+\frac{Q_m^2}{r^2}-\frac{2 \alpha Q_m^4}{5 r^6}-\frac{\Lambda _{\text{eff}}r^2}{3} \nonumber \\
&&+\nu\left(\frac{m}{r^2}+\frac{6 \alpha Q_m^4}{5 r^7}-\frac{Q_m^2}{r^3}-\frac{\Lambda _{\text{eff}}r}{3}\right)\Bigg],\label{d4}
\end{eqnarray}
and
\begin{eqnarray}
\dot{\omega}&=&-\frac{h^{2}}{(1-\omega^{2})}\Bigg[\nu  \left(-\frac{\Lambda_{\rm eff} }{3}-\frac{2 m}{r^3}-\frac{42 \alpha  Q_m^4}{5 r^8}+\frac{3 Q_m^2}{r^4}\right) \nonumber \\
&&~~~~~~~~~~+ \frac{2 m}{r^2}+\frac{12 \alpha  Q_m^4}{5 r^7}-\frac{2 Q_m^2}{r^3}-\frac{2 \Lambda_{\rm eff}  r}{3}\nonumber\\
&&~~~~~~~~~~ -a^{2} {\cal F} \Big(4\ln \sqrt{r(\nu + r)}\Big)_{,r} +(\ln {\cal F})_{,r}\Bigg].\label{d5} \nonumber\\
\end{eqnarray}
We set Eq. (\ref{d4}) and (\ref{d5}) equal to zero and solve simultaneously to find the critical points, which are given below
\begin{widetext}
\begin{eqnarray}
\omega_{c}^{2}&=&a_c^2= \frac{(\nu +r_{c})[5 r_{c}^5(6 m (\nu -r_{c})+\Lambda_{\rm eff}  r_{c}^3 (\nu +2 r_{c})]+18 \alpha  Q_m^4 (7 \nu -2 r_{c})+15 Q_m^2 r_{c}^4 (2 r_{c}-3 \nu )}{15 r_{c}^7[6 m+(\nu +2 r_{c})(\Lambda_{\rm eff}  r_{c} (\nu +r_{c})-2)]-6 \alpha  Q_m^4(-15 \nu ^2+2 r_{c}^2-5 \nu  r_{c})-15 Q_m^2 r_{c}^4 (\nu ^2+2 r_{c}^2-\nu  r_{c})},\label{d6}
\end{eqnarray}
where $r_{c}$, $a^{2}_{c}$ and $\omega_{c}$ represent the distance, speed of sound, and three-velocity of the fluid at the critical point also, we assume $a^{2}_{c}=k$.
Furthermore, we can use Eq. (\ref{c9}), to find the constant $C_{3}^{2}$ in terms of the critical points, we have
\begin{eqnarray}
C_{3}^{2}&=&\frac{1}{450 r_c^{12}(2 r_c+\nu)} n_{c}^2(r_c+\nu){}^3\Big[5 r_c^8 \Lambda _{\text{eff}}(r_c+\nu) +6 \alpha  Q_m^4 (r_c-3 \nu)+15 r_c^4 Q_m^2(\nu -r_c) -15 r_c^5(-2 m r_c+r_c^2+m \nu)\Big]    \nonumber\\
&&~~~~~~~~~ \times \Big[5 r_c^8 \Lambda _{\text{eff}}(2 r_c+\nu)+18 \alpha  Q_m^4(7 \nu -2 r_c) +15 r_c^4 Q_m^2(2 r_c-3 \nu)+30 m r_c^5(\nu -r_c)\Big].\label{d7}
\end{eqnarray}
By using Eqs. (\ref{c9}) and (\ref{d6}), the result would be
\begin{eqnarray}
\Big(\frac{n}{n_{c}}\Big)^{2}&=&-\Big(r^5(1-\omega ^2)(r_c+\nu)^{3}\Big(5 r_c^5(6 m(\nu -r_c)+\Lambda_{\rm eff}  r_c^3(2 r_c+\nu ))+18 \alpha  Q_m^4(7 \nu -2 r_c)+15 Q_m^2 r_c^4(2 r_c-3 \nu)\Big)
\nonumber\\
&&\Big(5 r_c^5(r_c(r_c(\Lambda_{\rm eff}  r_c(r_c+\nu)-3)+6 m)-3 m \nu)+6 \alpha  Q_m^4(r_c-3 \nu)+15 Q_m^2 r_c^4(\nu -r_c)\Big)\Big)\nonumber\\
&& \times \Big(30 \omega ^2 r_c^{12} (\nu +r)^2(2 r_c+\nu)(5 r^5(m (6 r-3 \nu )+r^2 (\Lambda_{\rm eff}  r (\nu +r)-3))+6 \alpha  Q_m^4 (r-3 \nu )+15 Q_m^2 r^4 (\nu -r))\Big)^{-1}, \label{d8}  \nonumber \\
\end{eqnarray}
If the solution of Eqs. (\ref{d4}) and (\ref{d5}), do not exist at the sonic point, one can define the reference points $(r_{0}, \omega_{0})$ through the phase depiction to arrive at \cite{b1}
\begin{eqnarray}
\Big(\frac{n}{n_{0}}\Big)^{2}&=&\Big(r_0^5 \omega ^2(\omega _0^2-1) (\nu +r)^2(5 r^5(m (6 r-3 \nu )+r^2 (\Lambda_{\rm eff}  r (\nu +r)-3))+6 \alpha  Q_m^4 (r-3 \nu )+15 Q_m^2 r^4 (\nu -r))\Big)^{-1}\nonumber\\ &&\Big(r^5(\omega ^2-1) \omega _0^2(\nu +r_0){}^2(-15 r_0^5(m \nu +Q_m^2)+30 m r_0^6-18 \alpha  \nu  Q_m^4+6 \alpha  Q_m^4 r_0+15 \nu  Q_m^2 r_0^4\nonumber\\
&&+5 \Lambda_{\rm eff}\nu r_0^8+5 \Lambda_{\rm eff}  r_0^9-15 r_0^7)\Big), \label{d9}
\end{eqnarray}
\end{widetext}
The previous equations will also be used to investigate the spherical accretion in several fluids.

\section{Applications to test fluids}

In this section, we will analyze different types of fluids that flow around the magnetically charged EH BH with scalar hair using the above results. Conclusively, we assume the polytropic and isothermal test fluids given in the following subsections.

\subsection{Isothermal test fluid}

The motion of fluid at a constant temperature is regarded as an isothermal flow. In other words, the speed of sound is constant in the process of accretion. This makes sure that at any radius, the speed of sound at the critical point is identical to the speed of sound of the accretion flow. Hence, our system is adiabatically operating in this situation; it is more conceivable that our fluid is flowing in an isothermal fashion. Because of this, we drive the general solution of isothermal EOS in this subsection with the form $p = ke$. Additionally, by using Eqs. (\ref{b5}) and (\ref{b9}), in EOS, we have $p = kF(n)$ and $G (n) = kF$, where $k$ denotes the state parameter and $0<k\leq 1$ \cite{b3}. The definition of the adiabatic sound speed is typically given as $a = dp/de$. Hence, when we compare the adiabatic sound speed to the equation of state, we get $a^{2} = k$. From Eq. (\ref{b9}), we attain
\begin{equation}
nF'(n)-F(n)=kF(n), \label{e1}
\end{equation}
which yields
\begin{equation}
e=F=\frac{e_{c}}{n_{c}^{k+1}}n^{k+1}. \label{e2}
\end{equation}
By using $p=ke$ and Eq. (\ref{e2}), in Eq. (\ref{a9}), we get
\begin{equation}
h=\frac{(k+1)e_{c}}{n_{c}}\left(\frac{n}{n_{c}}\right)^{k}. \label{e3}
\end{equation}
From Eq. (\ref{d8}), and Eq. (\ref{e3}) we have
\begin{widetext}
\begin{equation}
h^{2}\propto15^k \left[\frac{r^5 \left(\omega ^2-1\right)}{\omega ^2 (\nu +r)^2 \left(5 r^8 \Lambda _{\text{eff}} (\nu +r)+15 r^4 Q_m^2 (\nu -r)+6 \alpha  Q_m^4 (r-3 \nu )-15 r^5 \left(m \nu -2 m r+r^2\right)\right)}\right]{}^k, \label{e4}
\end{equation}
\end{widetext}
and
\begin{equation}
\mathcal{H}(r,\omega)=\frac{{\cal F}^{1-k}(r)}{(1-\omega^{2})^{1-k}\omega^{2k}(r (\nu +r))^{2k}}, \label{e5}
\end{equation}
where all constant components are integrated into the characterization of time $\bar{t}$ and Hamiltonian $\mathcal{H}$. Now, we will study how the fluid acts by using various values for the state parameter $k$.

\subsubsection{Solution for ultra-stiff fluid $(k=1$)}

Now, we assume the ultra-stiff fluid, which is obtained by setting $k = 1$ and $p=ke$ as the equation of state. This fluid has the characteristic that its energy density and isotropic pressure are both equal. The Hamiltonian Eq. (\ref{e5}), can be transformed into the form
\begin{equation}
\mathcal{H}=\frac{1}{r^2 \omega ^2 (\nu +r)^2}. \label{e6}
\end{equation}
From Eq. (\ref{e6}), it can be seen that the flow will be physical if $\mid \omega \mid < 1$. As a result, for the case of the ultra-stiff fluid, the minimum value of Hamiltonian (\ref{e6}), is $\mathcal{H}_{min}=\frac{1}{r^2(\nu +r)^2}$. Differentiating Eq. (\ref{e6}), with respect to $\omega$ and $r$, we get the following system of equation
\begin{equation}
\dot{\omega}=\frac{2 (\nu +2 r)}{r^3 \omega ^2 (\nu +r)^3},\label{e7}
\end{equation}
\begin{equation}
\dot{r}=-\frac{2}{r^2 \omega ^3 (\nu +r)^2}.\label{e8}
\end{equation}
\begin{figure}
\includegraphics[width=8.5cm]{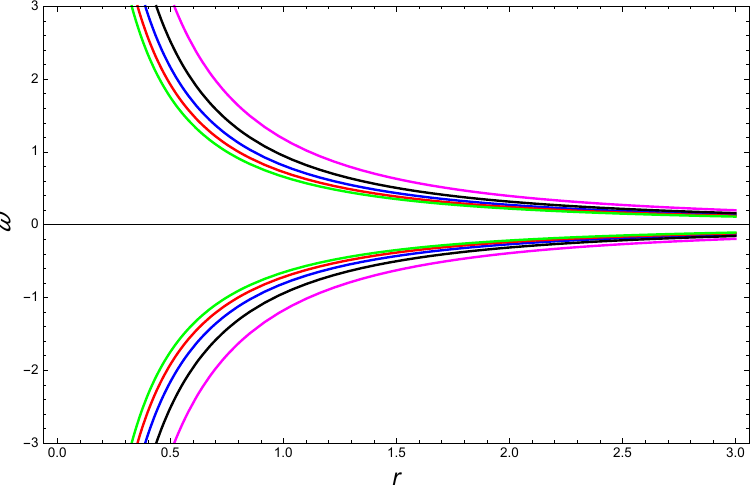}
\caption{The profile $\mathcal{H}$ (\ref{e7}) for an ultra-stiff fluid is as follows: for magnetically charged EEH BH with parameters $m=1$, $\nu=0.2$, $Q_m=0.2$, $\Lambda _{\text{eff}}=-1$, $\alpha =1$. The horizon can be located at $r_{h}\simeq 1.1723$.  The blue curve is an illustration of $\mathcal{H}= \mathcal{H}_{min}\simeq$ 0.38. The green and red curves refer to $\mathcal{H}> \mathcal{H}_{min}$ whereas the black and magenta curves represent the behavior for $\mathcal{H}< \mathcal{H}_{min}$ }.
\end{figure}
From Eqs. (\ref{e7}) and (\ref{e8}), it is clear that the above dynamical system does not have a critical point. The minimum value of $\mathcal{H}$ is $\mathcal{H}_{min}=r^{-2} (\nu +r)^{-2}$: $ \mathcal{H}>\mathcal{H}_{min}$. The curves in between the two magenta curves in \textbf{Fig. 2}, represent physical flows. Curves that lie on the upper half-plane with $\omega > 0$ show the outer flow of the fluid or the emission of particles, while curves on the lower half-plane with $\omega < 0$ show the fluid accretion.
\subsubsection{Solution for ultra-relativistic fluid ($k=\frac{1}{2}$)}
Now we study the ultra-relativistic fluid with the equation of state parameter $k = 1/2$. In this scenario, the energy
density of the fluids is greater than their isotropic pressure. {By assuming the value of state function $k=\frac{1}{4}$, Eq (\ref{d6}) gives us}
{\begin{multline}
N(r_{c})=0.5(-15 r^7((\nu +2 r)(r \Lambda _{\text{eff}} (\nu +r)-2)+6 m)\\+
6 \alpha  Q_m^4(-15 \nu ^2+2 r^2-5 \nu  r)+15 r^4 Q_m^2(\nu ^2+2 r^2-\nu r))+
(\nu +r)\\(5 r^5(r^3 \Lambda _{\text{eff}} (\nu +2 r)+6 m (\nu -r))+15 r^4 Q_m^2 (2 r-3 \nu )+
18 \alpha  Q_m^4\\ (7 \nu -2 r))=0, \label{f666}
\end{multline}}
{We can not find the roots
of the above polynomial analytically; its roots will be computed numerically. For this, we consider the following values
$\alpha =1$ $m=1$ $\nu =0.2$ $\Lambda _{\text{eff}}=-1$ $Q_m=2$ of the BH parameters and corresponding real roots are
$r_{c1} =0.692258$, $r_{c2} =1.16033$.}Â 
The Hamiltonian (\ref{e5}) has the
form
\begin{equation}
\mathcal{H}(r,\omega)=\frac{{\cal F}^{\frac{1}{2}}(r)}{(1-\omega^{2})^{\frac{1}{2}}\omega r (\nu +r)}. \label{e9}
\end{equation}
Further, the system of equations in Eqs. (\ref{d4}) and (\ref{d5}), takes the following form
\begin{widetext}
\begin{eqnarray}
\dot{r}&=&\frac{\sqrt{-5 r^8 \Lambda _{\text{eff}} (\nu +r)+15 r^4 Q_m^2 (r-\nu )-6 \alpha  Q_m^4 (r-3 \nu )+15 r^5(m \nu -2 m r+r^2)}}{\sqrt{15} r \sqrt{r^7}(1-\omega ^2)^{3/2} (\nu +r)}\nonumber\\
&&- \frac{\sqrt{-5 r^8 \Lambda _{\text{eff}} (\nu +r)+15 r^4 Q_m^2 (r-\nu )-6 \alpha  Q_m^4 (r-3 \nu )+15 r^5 (m \nu -2 m r+r^2)}}{\sqrt{15} r \sqrt{r^7} \omega ^2 \sqrt{1-\omega ^2} (\nu +r)},\label{f1} \\
\dot{\omega}&=&- \Big(5 r^8 \Lambda _{\text{eff}} (\nu +r) (\nu +2 r)+6 \alpha  Q_m^4(-27 \nu ^2+10 r^2-25 \nu  r)+15 r^4 Q_m^2(5 \nu ^2-6 r^2+3 \nu  r)-30 r^5(m(2 \nu ^2-5 r^2) \nonumber\\
&&+r^2 (\nu +2 r)) \Big)\Big(2 \sqrt{15} r^2 \sqrt{r^7} \omega  \sqrt{1-\omega ^2} (\nu +r)^2\Big(-5 r^8 \Lambda _{\text{eff}} (\nu +r)+15 r^4 Q_m^2 (r-\nu )-6 \alpha  Q_m^4 (r-3 \nu )\nonumber\\
&&+15 r^5(m \nu -2 m r+r^2) \Big)^{\frac{1}{2}}\Big)^{-1}.\label{f2}
\end{eqnarray}
\end{widetext}

\begin{figure}
\includegraphics[width=8.1 cm]{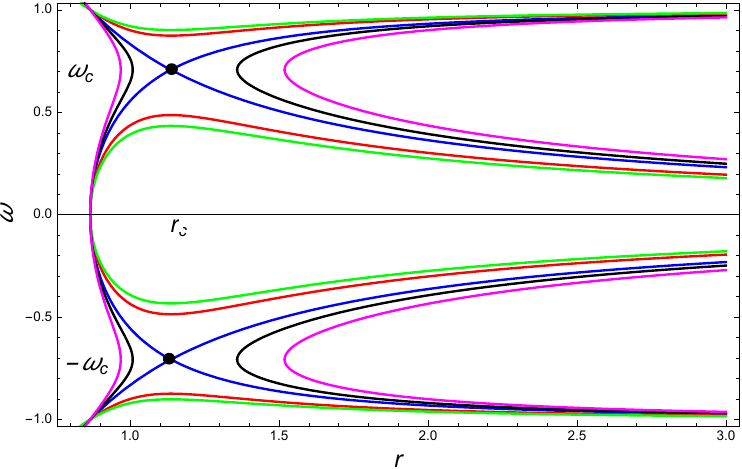}
\caption{The plot $\mathcal{H}$ (\ref{e9}) is for ultra-relativistic fluid with the BH parameters $m = 1$, $Q_{m} = 1$, $\nu = 0.2$, $ \Lambda _{\text{eff}} = -1$, $\alpha = 1$. The sonic (critical) points $(r_{c}, \omega_{c})$ are represented by black dots in a given plot. In the above figure, five plots are given, with the colors blue, red, green, and magenta corresponding with the given Hamiltonian values $\mathcal{H}=\mathcal{H}_{c}$, $\mathcal{H}=\mathcal{H}_{c}+0.2$, $\mathcal{H}=\mathcal{H}_{c}+0.4$, $\mathcal{H}=\mathcal{H}_{c}-0.2$, and $\mathcal{H}=\mathcal{H}_{c}- 0.3 $}.
\end{figure}

\begin{figure}
\includegraphics[width=8cm]{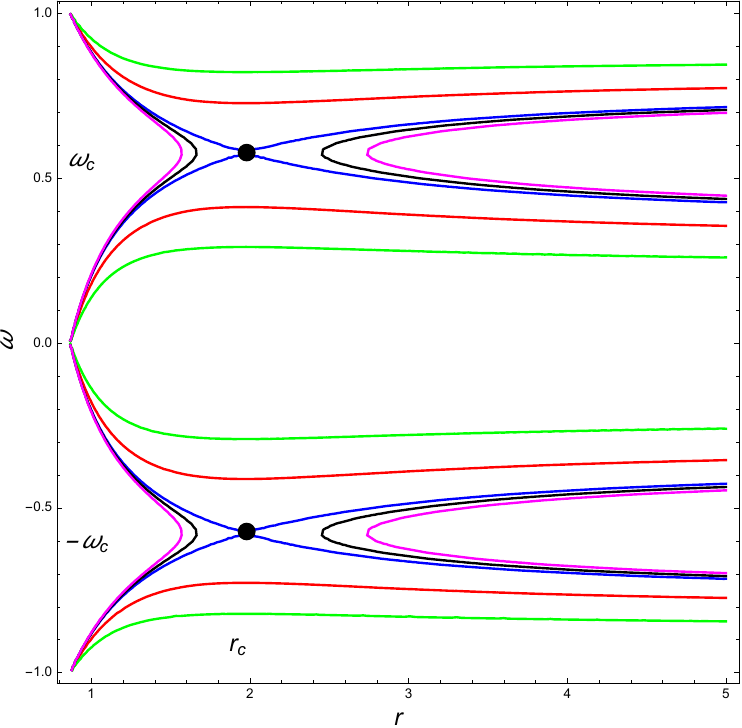}
\caption{The contour plot $\mathcal{H}$ (\ref{f3}) of radiation fluid with the BH parameters $m = 1$, $Q_{m} = 1$, $\nu = 0.2$, $ \Lambda _{\text{eff}}= -1$, $\alpha = 1$. The sonic (critical) points $(r_{c},\pm\omega_{c})$ are represented by black dots in a given plot. In the above-mentioned figure, five plots are given, with the colors blue, red, green, and magenta corresponding with the given Hamiltonian values $\mathcal{H}=\mathcal{H}_{c}$, $\mathcal{H}=\mathcal{H}_{c}+0.08263$, $\mathcal{H}=\mathcal{H}_{c}+0.28263$, $\mathcal{H}=\mathcal{H}_{c}-0.0037$, and $\mathcal{H}=\mathcal{H}_{c}- 0.01637$}.
\end{figure}

\begin{table*}[ht]
    \caption{The values of $\omega_{c}$, $r_{c}$ and $ \mathcal{H}_{c}$ at critical points with various values of BH parameters for sub-relativistic fluid are given} 
    \centering 
    \begin{tabular}{c c c c c c c c c c c c c c c c c c c c} 
      \hline 
         $\alpha=1$ ~~~~&$\nu=0.2$~~~ &$ \Lambda _{\text{eff}}=-1$ & $ $ & & & & & & & & &~~~~~~~~~~& &$ Q_{m}=2$ ~~~&$\nu=0.2$ ~~~&$ \Lambda _{\text{eff}}=-1$\\ [0.7ex] 
        \hline 
        \hline 
         $Q_{m}$\ &$ r_{c}$ &$ \omega_{c}$ & $ \mathcal{H}_{c}$ & & & & & & & & && $ \alpha$\ &$ r_{c}$ &$ \omega_{c}$ &$ \mathcal{H}_{c}$\\ [0.7ex] 
        \hline 
      2 &   2.4114392800453333 &0.5 & 1.55817 & &   & &   &  &  & & & & 1 &2.4114392800453333 & 0.5 & 1.55817\\
      2.1 & 2.514506523351315 & 0.5& 1.58055& & & &  & & & & & &  1.1 &2.4002507301569147 & 0.5& 1.55718\\
      2.2 & 2.610397375537204 & 0.5& 1.60123& &  & &   & & & & & &  1.2 & 2.3884127495453638 & 0.5& 1.55614\\
      2.3 & 2.7004847871562045 & 0.5& 1.62052& &  & &  & & & & & & 1.3 & 2.375826439742572 &0.5& 1.55508\\
      2.4 & 2.7857534742621994 & 0.5& 1.63865& &  & &   & & & & & &  1.4 & 2.3623660987961985 & 0.5& 1.55396\\
      \hline 
    \end{tabular}
    \label{table:nonlin}
\end{table*}

\begin{table*}[ht]
    \caption{The values of $\omega_{c}$, $r_{c}$ and $ \mathcal{H}_{c}$ at critical point with several values of BH parameters for radiation fluid are given.} 
    \centering 
    \begin{tabular}{c c c c c c c c c c c c c c c c c c c c} 
      \hline 
         $\alpha=1$ ~~~~&$Q_{m}=2$~~~ &$ \Lambda _{\text{eff}}=-1$ & $ $ & & & & & & & & &~~~~~~~~~~& &$ Q_{m}=2$ ~~~&$\nu=0.2$ ~~~&$\alpha=1$\\ [0.7ex] 
        \hline 
        \hline 
         $\nu$\ &$ r_{c}$ &$ \omega_{c}$ & $ \mathcal{H}_{c}$ & & & & & & & & && $\Lambda _{\text{eff}}$\ &$ r_{c}$ &$ \omega_{c}$ &$ \mathcal{H}_{c}$\\ [0.7ex] 
        \hline 
      0.1  &2.4616 & 0.5 & 1.55853& &  & & & & & & & & -1.4 &2.13062 & 0.5 & 1.87166\\
      0.11 & 2.45658 & 0.5&1.55851& & & &  & & & & & &  -1.3 &2.19145 & 0.5& 1.79725\\
      0.12 & 2.45155 & 0.5& 1.55848& &  & &   & & & & & &  -1.2 & 2.25751& 0.5& 1.72039\\
      0.13 &2.44654 & 0.5& 1.55845& &  & &  & & & & & & -1.1 & 2.330217 & 0.5& 1.6408\\
      0.14 & 2.44152 & 0.5& 1.55845& &  & &   & & & & & &  -1 & 2.4114 & 0.5& 1.55817\\
      \hline 
    \end{tabular}
    \label{table:nonlin}
\end{table*}

Also, we can find critical points for ultra-relativistic fluid, when the RHS of Eqs. (\ref{f1}) and (\ref{f2}), vanish. By considering the BH parameters $m = 1$, $Q_{m} = 1$, $\alpha = 1$, $\Lambda _{\text{eff}} = -1$ and $\nu = 0.2$, we can obtain the physical critical points $(r_{c}, \pm\omega_{c})$, which are $(1.13718, -0.707107)$ for the outflow of the fluid and $(1.13718, 0.707107)$ for fluid accretion, respectively. By substituting these critical points into Eq. (\ref{e9}), we obtain the critical Hamiltonian $\mathcal{H}_{c}=0.8849$. The critical points $r_{c}$, $\omega_{c}$, and $\mathcal{H}_{c}$ for various values of the BH parameter are listed in \textbf{tables I and II}. In \textbf{Fig. 3}, we represent the physical behavior of an ultra-relativistic fluid by numerous curves with BH parameters $m = 1$, $Q_{m} = 1$, $\alpha = 1$, $\Lambda _{\text{eff}} = -1$ and $\nu = 0.2$. From \textbf{Fig. 3}, we observe that the critical points $(r_{c}, \omega_{c})$ and $(r_{c}, -\omega_{c})$ are the saddle points of the given dynamical system. In \textbf{Fig. 3}, the red curve (with $\mathcal{H}=\mathcal{H}_{c}+0.2$ branches) and the green curve (with $\mathcal{H}= \mathcal{H}_{c}+0.4$ branches) show purely supersonic outer flow ($\omega>\omega_{c}$ branches), supersonic accretion $(\omega<-\omega_{c})$, or purely subsonic accretion followed by subsonic outer flow $(-\omega_{c}<\omega<\omega_{c})$, respectively. The black (with $\mathcal{H}=\mathcal{H}_{c}$-0.2 branches) and magenta (with $\mathcal{H}=\mathcal{H}_{c}-0.3$ branches) curves represent no physical behavior.

\begin{table*}[ht]
    \caption{The values of $\omega_{c}$, $r_{c}$ and $ \mathcal{H}_{c}$ at critical points with various values of BH parameters for ultra-relativistic fluid are given} 
    \centering 
    \begin{tabular}{c c c c c c c c c c c c c c c c c c c c} 
      \hline 
         $\alpha=1$ ~~~~&$\nu=0.2$~~~ &$ \Lambda _{\text{eff}}=-1$ & $ $ & & & & & & & & &~~~~~~~~~~& &$ Q_{m}=1$ ~~~&$\nu=0.2$ ~~~&$ \Lambda _{\text{eff}}=-1$\\ [0.7ex] 
        \hline 
        \hline 
         $Q_{m}$\ &$ r_{c}$ &$ \omega_{c}$ & $ \mathcal{H}_{c}$ & & & & & & & & && $ \alpha$\ &$ r_{c}$ &$ \omega_{c}$ &$ \mathcal{H}_{c}$\\ [0.7ex] 
        \hline 
      1 &1.1372 & 0.7071 & 0.8849 & &   & &   &  &  & & & & 1 &1.1372 & 0.7071 & 0.8849\\
      1.1 & 1.0697 & 0.7071& 0.9833& & & &  & & & & & &  1.1 &1.1516, & 0.7071& 0.876828\\
      1.2 & 1.0318 & 0.7071& 1.08557& &  & &   & & & & & &  1.2 & 1.1647, & 0.7071& 0.869434\\
      1.3 & 1.0192 & 0.7071& 1.1749& &  & &  & & & & & & 1.3 & 1.17693, & 0.7071& 0.862649\\
      1.4 & 1.0235 & 0.7071& 1.2438& &  & &   & & & & & &  1.4 & 1.1882, & 0.7071& 0.85638\\
      \hline 
    \end{tabular}
    \label{table:nonlin}
\end{table*}

\begin{table*}[ht]
    \caption{The values of $\omega_{c}$, $r_{c}$ and $ \mathcal{H}_{c}$ at a critical point with altered values of BH parameters for ultra-relativistic fluid are given} 
    \centering 
    \begin{tabular}{c c c c c c c c c c c c c c c c c c c c} 
      \hline 
         $\alpha=1$ ~~~~&$Q_{m}=1$~~~ &$ \Lambda _{\text{eff}}=-1$ & $ $ & & & & & & & & &~~~~~~~~~~& &$ Q_{m}=1$ ~~~&$\nu=0.2$ ~~~&$\alpha=1$\\ [0.7ex] 
        \hline 
        \hline 
         $\nu$\ &$ r_{c}$ &$ \omega_{c}$ & $ \mathcal{H}_{c}$ & & & & & & & & && $\Lambda _{\text{eff}}$\ &$ r_{c}$ &$ \omega_{c}$ &$ \mathcal{H}_{c}$\\ [0.7ex] 
        \hline 
      0.1  &1.21881  & 0.7071 & 0.86931& &  & & & & & & & & -1.4 &1.07336 & 0.7071 & 1.0743\\
      0.11 & 1.2122 & 0.7071& 0.87014& & & &  & & & & & &  -1.3 &1.08747 & 0.7071& 1.02847\\
      0.12 & 1.2055 & 0.7071& 0.87109& &  & &   & & & & & &  -1.2 &1.10269 & 0.7071& 0.98169\\
      0.13 & 1.1984 & 0.7071& 0.87217& &  & &  & & & & & & -1.1 & 1.11918 & 0.7071& 0.93388\\
      0.14 & 1.1911 & 0.7071& 0.87339& &  & &   & & & & & &  -1 & 1.13718 & 0.7071& 0.88495\\
      \hline 
    \end{tabular}
    \label{table:nonlin}
\end{table*}

In \textbf{Fig. 3}, the blue curves represent the fascinating solution of the fluid and reveal the transonic behavior of the fluid outside the BH horizon. For $\omega<0 $, the curves that traverse across the sonic point $(r_{c}, \omega_{c})$. One solution begins at spatial infinity with subsonic flow and proceeds to supersonic flow after crossing the sonic point. This solution refers to the standard non-relativistic accretion investigated by Bondi \cite{c1}. In accordance with the investigation \cite{b1}, the alternative solution that proceeds at spatial infinity with the supersonic flow but transforms to subsonic after crossing the sonic point is unstable, so it is extremely challenging to find these behaviors. If $\omega > 0$, there are two possible solutions. One solution is detailed in \cite{c1} for non-relativistic accretion, which correlates to the transonic solution of the stellar wind, initiating flow at the horizon with supersonic flow and switching to subsonic flow after crossing the sonic point, while the remaining solution is identical to the case $\omega<0$, which is hard to attain and unstable \cite{b1}.
Generally, various Hamiltonian values correspond to various initial states of the dynamical system. If the ultra-relativistic fluid has a transonic solution, then the Hamiltonian can be examined at the sonic point. The Hamiltonian with distinct values from the transonic one cannot represent any transonic flow solutions. For instance, the magenta curve displays the subcritical flow of the fluid since flows will not reach the critical point. In reality, the solutions have a turning point, which is the closest point that such fluids can reach before rebounding back or turning around infinity. Similarly, the black curves can be explained. Moreover, super-critical flows can be seen in the green and red curves. Although fluids do not reach critical points, their velocities are already higher than the permissible critical value. Such flows finally enter the BH horizon. It is also worth noting that the same evaluation applies to other fluids, such as radiation, sub-relativistic and polytropic fluids.
\subsubsection{Solution for radiation fluid ($k=\frac{1}{3}$)}
The fluid that absorbs radiation emitted from the BH is said to be radiation fluid.
{By considering the value of state function $k=\frac{1}{3}$, Eq (\ref{d6}) gives us}
{\begin{multline}
N(r_{c})=1/3(-15 r^7((\nu +2 r)(r \Lambda _{\text{eff}} (\nu +r)-2)+6 m)\\+
6 \alpha  Q_m^4(-15 \nu ^2+2 r^2-5 \nu  r)+15 r^4 Q_m^2(\nu ^2+2 r^2-\nu r))+
(\nu +r)\\(5 r^5(r^3 \Lambda _{\text{eff}} (\nu +2 r)+6 m (\nu -r))+15 r^4 Q_m^2 (2 r-3 \nu )+
18 \alpha  Q_m^4\\ (7 \nu -2 r))=0, \label{f666}
\end{multline}}
{We can not find the roots
of the above polynomial analytically; its roots will be computed numerically. For this, we consider the following values
$\alpha =1$ $m=1$ $\nu =0.2$ $\Lambda _{\text{eff}}=-1$ $Q_m=2$ of the BH parameters and corresponding real roots are
$r_{c1} =0.702594$, $r_{c2} =1.27678$.}
The Hamiltonian (\ref{e5}), becomes
\begin{equation}
\mathcal{H}(r,\omega)=\frac{{\cal F}^{2/3}(r)}{(1-\omega^{2})^{2/3}\omega^{2/3}(r (\nu +r))^{2/3}}, \label{f3}
\end{equation}
and the above system (\ref{d4}, \ref{d5}) reduce to
\begin{widetext}
\begin{eqnarray}
\dot{\omega}&=&\frac{2 (\nu +2 r)(-5 r^8 \Lambda _{\text{eff}} (\nu +r)+15 r^4 Q_m^2 (r-\nu )-6 \alpha  Q_m^4 (r-3 \nu )+15 r^5(m \nu -2 m r+r^2)){}^{2/3}}{3\ 15^{2/3}(r^7)^{2/3} \omega ^{2/3}(1-\omega ^2)^{2/3} (r (\nu +r))^{5/3}}\nonumber\\
&&\frac{14 r^6(-5 r^8 \Lambda _{\text{eff}} (\nu +r)+15 r^4 Q_m^2 (r-\nu )-6 \alpha  Q_m^4 (r-3 \nu )+15 r^5(m \nu -2 m r+r^2)){}^{2/3}}{3\ 15^{2/3}(r^7)^{5/3} \omega ^{2/3}(1-\omega ^2)^{2/3} (r (\nu +r))^{2/3}}\nonumber\\
&&-\Big(2(-5 r^8 \Lambda _{\text{eff}}-40 r^7 \Lambda _{\text{eff}} (\nu +r)-6 \alpha  Q_m^4+15 r^4 Q_m^2+60 r^3 Q_m^2 (r-\nu )+15 r^5 (2 r-2 m)+\nonumber\\
&&75 r^4(m \nu -2 m r+r^2))\Big)\Big(3\ 15^{2/3}(r^7)^{2/3} \omega ^{2/3}(1-\omega ^2)^{2/3} (r (\nu +r))^{2/3}(-5 r^8 \Lambda _{\text{eff}} (\nu +r)+15 r^4 Q_m^2 (r-\nu )\nonumber\\
&&-6 \alpha  Q_m^4 (r-3 \nu )+15 r^5(m \nu -2 m r+r^2)\Big)^{\frac{1}{3}}\Big)^{-1},\label{f4}
\end{eqnarray}
\begin{eqnarray}
\dot{r}&=&\frac{4 \sqrt[3]{\omega } \left(-5 r^8 \Lambda _{\text{eff}} (\nu +r)+15 r^4 Q_m^2 (r-\nu )-6 \alpha  Q_m^4 (r-3 \nu )+15 r^5 \left(m \nu -2 m r+r^2\right)\right){}^{2/3}}{3\ 15^{2/3} \left(r^7\right)^{2/3} \left(1-\omega ^2\right)^{5/3} (r (\nu +r))^{2/3}}\nonumber\\&&-\frac{2 \left(-5 r^8 \Lambda _{\text{eff}} (\nu +r)+15 r^4 Q_m^2 (r-\nu )-6 \alpha  Q_m^4 (r-3 \nu )+15 r^5 \left(m \nu -2 m r+r^2\right)\right){}^{2/3}}{3\ 15^{2/3} \left(r^7\right)^{2/3} \omega ^{5/3} \left(1-\omega ^2\right)^{2/3} (r (\nu +r))^{2/3}}. \label{f5}
\end{eqnarray}
\end{widetext}
We can determine the critical points of the above dynamical system by setting the right-hand side of Eqs. (\ref{f4}) and (\ref{f5}), equal to zero and then solving for $r$ and $\omega$. Moreover, in \textbf{tables III and IV}, the critical values $r_{c}$, $\omega_{c}$, and $\mathcal{H}_{c}$ are shown for the different values of the BH parameters. In \textbf{Fig. 4}, several curves illustrate the physical behavior of radiation fluid with BH parameters $m = 1$, $Q_{m} = 1$, $\alpha = 1$, $\Lambda _{\text{eff}} = -1$ and $\nu = 0.2$. From \textbf{Fig. 4}, it is clear that the critical points $(r_{c}, -\omega_{c})$ and $(r_{c}, \omega_{c})$ seem to be saddle points of the above dynamical system. Furthermore, from \textbf{Fig. 4}, it is easy to observe that the motion of the radiation fluid $(k = \frac{1}{3})$ is identical to that of the ultra-relativistic fluid, which is depicted in \textbf{Fig. 3}. Here, red and green contours represent supersonic flows when $\omega < -\omega_{c}$ or $\omega >\omega_{c}$, and subsonic flows when $-\omega <\omega_{c}< \omega$, while the same behavior is observed in \textbf{Fig. 3}, for ultra-stiff fluid. The blue curves illustrate the transonic solutions. For $\omega < 0$, one of the blue curves (which begins at spatial infinity with subsonic flow followed by supersonic after crossing the sonic point $(r_{c}, \omega_{c})$ shows the usual transonic accretion, while the other blue curve demonstrates the solution is unstable. For $\omega > 0$, one blue curve represents the transonic outer flow of wind, while the other indicates that the flow is unstable, analogous to ultra-relativistic fluid. The magenta and black curves demonstrate the non-physical solution.
\subsubsection{Sub-relativistic fluid ($k=\frac{1}{4}$):  Separatrix
heteroclinic flows}
We assume that the state equation for a sub-relativistic fluid is $p = e/4$, which means that the energy density is higher
than its isotropic pressure. {By considering the value of state function $k=\frac{1}{4}$, Eq (\ref{d6}) gives us}
{\begin{multline}
N(r_{c})=0.25(-15 r^7((\nu +2 r)(r \Lambda _{\text{eff}} (\nu +r)-2)+6 m)\\+
6 \alpha  Q_m^4(-15 \nu ^2+2 r^2-5 \nu  r)+15 r^4 Q_m^2(\nu ^2+2 r^2-\nu r))+
(\nu +r)\\(5 r^5(r^3 \Lambda _{\text{eff}} (\nu +2 r)+6 m (\nu -r))+15 r^4 Q_m^2 (2 r-3 \nu )+
18 \alpha  Q_m^4\\ (7 \nu -2 r))=0, \label{f666}
\end{multline}}
{To investigate heteroclinic flow, we compute the roots of the above polynomial. But we are unable to determine the roots
of the above polynomial analytically; its roots will be computed numerically. For this, we consider the following values
$\alpha =1$ $m=1$ $\nu =0.2$ $\Lambda _{\text{eff}}=-1$ $Q_m=2$ of the BH parameters and corresponding real roots are
$r_{c1} =1.25842$, $r_{c2} =2.6262$.}Â 
The Hamiltonian (\ref{e5}) becomes
\begin{equation}
\mathcal{H}(r,\omega)=\frac{{\cal F}^{3/4}(r)}{(1-\omega^{2})^{3/4}\omega^{1/2}(r (\nu +r))^{1/2}}, \label{f6}
\end{equation}
and dynamical two-dimensional system (\ref{d4}, \ref{d5}) read as
\begin{widetext}
\begin{eqnarray}
\dot{\omega}&=&\frac{(\nu +2 r)(-5 r^8 \Lambda _{\text{eff}} (\nu +r)+15 r^4 Q_m^2 (r-\nu )-6 \alpha  Q_m^4 (r-3 \nu )+15 r^5(m \nu -2 m r+r^2)){}^{3/4}}{2\ 15^{3/4}(r^7)^{3/4} \sqrt{\omega }(1-\omega ^2)^{3/4} (r (\nu +r))^{3/2}}\nonumber\\
&&+\frac{7 \sqrt[4]{3} r^6(-5 r^8 \Lambda _{\text{eff}} (\nu +r)+15 r^4 Q_m^2 (r-\nu )-6 \alpha  Q_m^4 (r-3 \nu )+15 r^5(m \nu -2 m r+r^2)){}^{3/4}}{4\ 5^{3/4}(r^7)^{7/4} \sqrt{\omega }(1-\omega ^2)^{3/4} \sqrt{r (\nu +r)}}\nonumber\\
&&-\Big(\sqrt[4]{3}(-5 r^8 \Lambda _{\text{eff}}-40 r^7 \Lambda _{\text{eff}} (\nu +r)-6 \alpha  Q_m^4+15 r^4 Q_m^2+60 r^3 Q_m^2 (r-\nu )+15 r^5 (2 r-2 m)\nonumber\\
&&+75 r^4(m \nu -2 m r+r^2))\Big)\Big(4\ 5^{3/4}(r^7)^{3/4} \sqrt{\omega }(1-\omega ^2)^{3/4} \sqrt{r (\nu +r)}(-5 r^8 \Lambda _{\text{eff}} (\nu +r)+15 r^4 Q_m^2 (r-\nu )\nonumber\\
&&-6 \alpha  Q_m^4 (r-3 \nu )+15 r^5(m \nu -2 m r+r^2))^{\frac{1}{4}}\Big)^{-1},\label{f7}
\end{eqnarray}
\begin{eqnarray}
\dot{r}&=&\frac{\sqrt[4]{3} \sqrt{\omega }(-5 r^8 \Lambda _{\text{eff}} (\nu +r)+15 r^4 Q_m^2 (r-\nu )-6 \alpha  Q_m^4 (r-3 \nu )+15 r^5(m \nu -2 m r+r^{2})){}^{3/4}}{2\ 5^{3/4}(r^7)^{3/4}(1-\omega ^2)^{7/4} \sqrt{r (\nu +r)}}\nonumber\\
&&-\frac{(-5 r^8 \Lambda _{\text{eff}} (\nu +r)+15 r^4 Q_m^2 (r-\nu )-6 \alpha  Q_m^4 (r-3 \nu )+15 r^5(m \nu -2 m r+r^2)){}^{3/4}}{2\ 15^{3/4}(r^7)^{3/4} \omega ^{3/2}(1-\omega ^2)^{3/4} \sqrt{r (\nu +r)}}. \label{f8}
\end{eqnarray}
\end{widetext}
\textbf{Figure 5} shows the phase space profiles for the sub-relativistic fluid. From \textbf{Fig. 5}, it can be seen
that the motion of the sub-relativistic fluid $(k =\frac{1}{4})$ is identical to the motion of the radiation
$(k=\frac{1}{3})$ and ultra-relativistic fluids
$(k=\frac{1}{2})$. The green and red curves are purely supersonic outer flows for $\omega > \omega_{c}$, while they
represent supersonic accretions for $\omega<- \omega_{c}$. Also these curves display subsonic flows when
$-\omega_{c}< \omega <\omega_{c}$. The blue curves in \textbf{Fig. 5}, are fascinating because they depict
the transonic solution of outer flow for $\omega > 0$ and the spherical accretion for $\omega< 0$. The magenta
and black curves are un-physical solutions, just like the radiation fluid and ultra-relativistic fluid.
In accordance with the previous two cases, we provide the values of $r_{c}$, $\omega_{c}$, and $\mathcal{H}_{c}$ for such
a dynamical system in \textbf{tables V and VI} with various values of the BH parameters.
{\textbf{Figure 6} is the contour of $\mathcal{H}$ (\ref{e5}) in $(r, \omega)$ plane. From \textbf{Fig. 6}, we observe that
$(r_{c1},\pm \omega_{c})$ are the two saddle points. We assume $(r_{rm},-\omega_{c})$, $(r_{rm},-\omega_{c})$ are the two
rightmost points in the lower and upper branches of the given plot, respectively. If we consider $\frac{d\omega}{dr}$
continuous, as the fluid passes through the saddle point, the processes of accretion will begin from $(r_{rm},-\omega_{c})$
on the Black curve in the lower curve of \textbf{plot 6}. If the mobility is subsonic, it flows into the upper branch of the
lower plot, passes through saddle point $(r_{c1}, \omega_{c})$, and then crosses the horizon. Alternatively, if the motion
is supersonic, it travels across the lower branch of the lower plot, eventually throughout the saddle point $(r_{c1}, \omega_{c})$,
when more till $\omega$ vanishes as the fluid nears the horizon. From there, the fluid travels along the upper branch of
the upper plot while traveling at supersonic speeds until the point $(r_{rm} ,\omega_{c})$ Â of the upper plot. First,
it may be demonstrated that such motion is unstable using considerations similar to those provided in the case when k = 1/2.
Secondly, the movement might become periodic, but this is extremely difficult to achieve.}
\begin{table*}[ht]
    \caption{The values of $\omega_{c}$, $r_{c}$ and $ \mathcal{H}_{c}$ at critical point with various values of BH parameters for radiation fluid are given} 
    \centering 
    \begin{tabular}{c c c c c c c c c c c c c c c c c c c c} 
      \hline 
         $\alpha=1$ ~~~~&$\nu=0.2$~~~ &$ \Lambda _{\text{eff}}=-1$ & $ $ & & & & & & & & &~~~~~~~~~~& &$ Q_{m}=1$ ~~~&$\nu=0.2$ ~~~&$ \Lambda _{\text{eff}}=-1$\\ [0.7ex] 
        \hline 
        \hline 
         $Q_{m}$\ &$ r_{c}$ &$ \omega_{c}$ & $ \mathcal{H}_{c}$ & & & & & & & & && $ \alpha$\ &$ r_{c}$ &$ \omega_{c}$ &$ \mathcal{H}_{c}$\\ [0.7ex] 
        \hline 
      1 &1.9698091867345635 & 0.57735 & 1.01719 & &   & &   &  &  & & & & 1 &1.9698091867345635 & 0.57735 & 1.01719\\
      1.1 & 1.6443248116131028 & 0.57735& 1.0427& & & &  & & & & & &  1.1 &1.9758777367178266 & 0.57735& 1.017\\
      1.2 & 1.3559118143943665 & 0.57735& 1.09489& &  & &   & & & & & &  1.2 & 1.981771727185524 & 0.57735& 1.01682\\
      1.3 & 1.2369938908525624 & 0.57735& 1.172& &  & &  & & & & & & 1.3 & 1.9875025608549501 & 0.57735& 1.01664\\
      1.4 & 1.1963945875650448 & 0.57735& 1.25507& &  & &   & & & & & &  1.4 & 1.993080497767953 & 0.57735&1.01646\\
      \hline 
    \end{tabular}
    \label{table:nonlin}
\end{table*}

\begin{table*}[ht]
    \caption{The values of $\omega_{c}$, $r_{c}$ and $ \mathcal{H}_{c}$ at critical point with several values of BH parameters for radiation fluid are given.} 
    \centering 
    \begin{tabular}{c c c c c c c c c c c c c c c c c c c c} 
      \hline 
         $\alpha=1$ ~~~~&$Q_{m}=1$~~~ &$ \Lambda _{\text{eff}}=-1$ & $ $ & & & & & & & & &~~~~~~~~~~& &$ Q_{m}=1$ ~~~&$\nu=0.2$ ~~~&$\alpha=1$\\ [0.7ex] 
        \hline 
        \hline 
         $\nu$\ &$ r_{c}$ &$ \omega_{c}$ & $ \mathcal{H}_{c}$ & & & & & & & & && $\Lambda _{\text{eff}}$\ &$ r_{c}$ &$ \omega_{c}$ &$ \mathcal{H}_{c}$\\ [0.7ex] 
        \hline 
      0.1  &2.027092265508927 & 0.577351 & 1.01672& &  & & & & & & & & -1.4 &1.9698091867345637 & 0.7071 & 1.23491\\
      0.11 & 2.021610540472511 & 0.577351& 1.01675& & & &  & & & & & &  -1.3 &1.9698091867345657 & 0.7071& 1.18239\\
      0.12 & 2.0160775177538173 & 0.577351& 1.01679& &  & &   & & & & & &  -1.2 & 1.9698091867345555& 0.7071& 1.12868\\
      0.13 & 2.0104920214953585 & 0.577351& 1.01683& &  & &  & & & & & & -1.1 & 1.9698091867345604 & 0.577351& 1.07366\\
      0.14 & 2.004852805359424 & 0.577351& 1.01687& &  & &   & & & & & &  -1 & 1.9698091867345635 & 0.577351& 1.01719\\
      \hline 
    \end{tabular}
    \label{table:nonlin}
\end{table*}
\begin{figure}
\includegraphics[width=8cm]{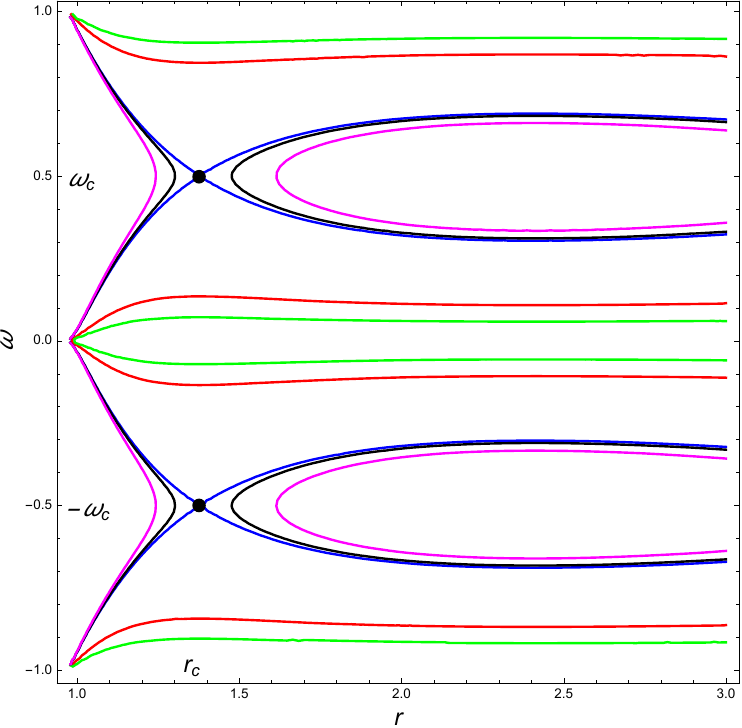}
\caption{Contour plot of $\mathcal{H}$ (\ref{f3}) is for sub-relativistic fluid with the BH parameters
$m = 1$, $Q_{m} = 2$, $\nu = 0.2$, $ \Lambda _{\text{eff}}= -1$, $\alpha = 1$. The sonic (critical) points
$(r_{c},\pm\omega_{c})$ are displayed by black dots in a given plot. In the above-mentioned figure, five plots are
given, with the colors blue, red, green, and magenta corresponding to the given Hamiltonian values
$\mathcal{H}=\mathcal{H}_{c}$, $\mathcal{H}=\mathcal{H}_{c}+1$, $\mathcal{H}=\mathcal{H}_{c}+2$,
$\mathcal{H}=\mathcal{H}_{c}-1.014$, and $\mathcal{H}=\mathcal{H}_{c}- 0.054$}.
\end{figure}
\begin{figure}
\includegraphics[width=8cm]{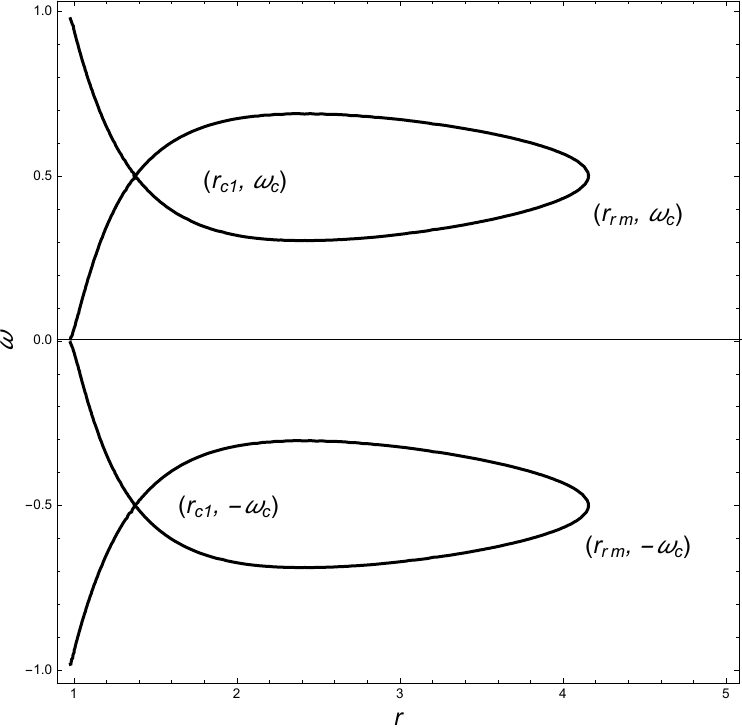}
{\caption{Contour plot of $\mathcal{H}$ (\ref{f3}) with the BH parameters $m = 1$, $Q_{m} = 2$, $\nu = 0.2$,
$ \Lambda _{\text{eff}}= -1$, $\alpha = 1$. In the given plot the heteroclinic solution curve saddle or
critical points $(r_{c1},\pm\omega_{c})$ for $\mathcal{H}(r_{c1},\omega_{c})$=$\mathcal{H}(r_{c1},-\omega_{c})\simeq 1.73384.$}}
\end{figure}

\subsection{Polytropic test fluid}

The equation of state for polytropic test fluid is characterized by
\begin{equation}
p=G(n)=Kn^{\gamma}, \label{f9}
\end{equation}
here, $\gamma$ and $K$ are taken to be constants. The constraint $\gamma>1$ is typically used when working with ordinary matter. We acquire the subsequent equations for the specific enthalpy using \cite{a9}
\begin{equation}
h=m+\frac{K\gamma n^{\gamma-1}}{\gamma-1}, \label{g1}
\end{equation}
in which the baryonic mass $m$ is known integration constant. The speed of sound is determined by
\begin{equation}
a^{2}=\frac{(\gamma-1)Y}{m(\gamma-1)+Y}\,\,\,\,\,\,\ Y=K\gamma n^{\gamma-1}. \label{g2}
\end{equation}
From Eqs. (\ref{d8}) and (\ref{g2}), we have
\begin{widetext}
\begin{equation}
h=m\Bigg\{1+Z\Bigg[\frac{1-\omega^{2}}{(r (\nu +r))^{2}(1-\frac{2 m}{r}+\frac{Q_m^2}{r^2}-\frac{2 \alpha Q_m^4}{5 r^6}-\frac{\Lambda _{\text{eff}}r^2}{3})+\nu(\frac{m}{r^2}+\frac{6 \alpha Q_m^4}{5 r^7}-\frac{Q_m^2}{r^3}-\frac{\Lambda _{\text{eff}}r}{3})}\Bigg]^{\frac{\gamma-1}{2}}\Bigg\} ,\label{g3}
\end{equation}
where
\begin{eqnarray}
Z&=&\frac{K\gamma}{m(\gamma-1)450 r_c^{12}(2 r_c+\nu)}\Big[n^2(r_c+\nu){}^3(5 r_c^8 \Lambda _{\text{eff}}(r_c+\nu)+6 \alpha  Q_m^4(r_c-3 \nu)+15 r_c^4 Q_m^2(\nu -r_c)\nonumber\\
&&-15 r_c^5(-2 m r_c+r_c^2+m \nu))(5 r_c^8 \Lambda _{\text{eff}}(2 r_c+\nu)+18 \alpha  Q_m^4(7 \nu -2 r_c)\nonumber\\
&&+15 r_c^4 Q_m^2(2 r_c-3 \nu)+30 m r_c^5(\nu -r_c))\Big]^{\frac{\gamma-1}{2}}.\label{g4}
\end{eqnarray}
From Eq. (\ref{g3}), $Z$ is positive constant.
By using Eqs. (\ref{g3}) and (\ref{d2}), we obtain
\begin{eqnarray}
\mathcal{H}&=&\frac{1}{1-\omega^{2}}\Big((1-\frac{2 m}{r}+\frac{Q_m^2}{r^2}-\frac{2 \alpha Q_m^4}{5 r^6}-\frac{\Lambda _{\text{eff}}r^2}{3})+\nu(\frac{m}{r^2}+\frac{6 \alpha Q_m^4}{5 r^7}-\frac{Q_m^2}{r^3}-\frac{\Lambda _{\text{eff}}r}{3})\Big)\nonumber\\ &&\Big[\Big(1+Z\Big(\frac{1-\omega^{2}}{(r (\nu +r))^{2}\Big((1-\frac{2 m}{r}+\frac{Q_m^2}{r^2}-\frac{2 \alpha Q_m^4}{5 r^6}-\frac{\Lambda _{\text{eff}}r^2}{3})+\nu(\frac{m}{r^2}+\frac{6 \alpha Q_m^4}{5 r^7}-\frac{Q_m^2}{r^3}-\frac{\Lambda _{\text{eff}}r}{3})\Big)}\Big)^{\frac{\gamma-1}{2}}\Big)\Big]^{2}, \label{g5}
\end{eqnarray}
\end{widetext}
here, $m^{2}$ is consumed in the re-definition of $(\bar{t}, \mathcal{H})$. As the Hamiltonian is constant on the solution curve, so no global solutions exist.
Using the process described in \cite{a9, b1, b5}, the following relationship can be obtained
\begin{widetext}
\begin{eqnarray}
a_{c}^{2}&=&\Big[\Big((\nu +r_{c})^2 \omega_{c}^{2}\Big(-5 r_{c}^5\Big(m (6r_{c}-3 \nu )+r_{c}^2 (\Lambda r_{c} (\nu +r_{c})-3)\Big)-6\alpha Q_m^4 (r_{c}-3 \nu )-15 Q_m^2r^4(\nu -r_{c})\Big)\Big)^{\frac{\gamma-1}{2}}\Big]^{-1}\nonumber\\
&& ~~~~~ \times \Big((\gamma-1-\omega_{c}^{2})\Big((1-\omega_{c}^{2})15 r_{c}^5\Big)^{\frac{\gamma-1}{2}}\Big),\label{g7} \\
\omega_{c}^{2}&=&\Big((\nu +r_{c})(5 r_{c}^5(6 m (\nu -r_{c})+\Lambda_{\rm eff}  r_{c}^3 (\nu +2 r_{c}))+18 \alpha  Q_m^4 (7 \nu -2 r_{c})+15 Q_m^2 r_{c}^4 (2 r_{c}-3 \nu ))\Big) \times \nonumber\\
&&\Big(15 r_{c}^7(6 m+(\nu +2 r_{c})(\Lambda_{\rm eff}  r_{c} (\nu +r_{c})-2))-6 \alpha  Q_m^4(-15 \nu ^2+2 r_{c}^2-5 \nu  r_{c})-15 Q_m^2 r_{c}^4 (\nu ^2+2 r_{c}^2-\nu  r_{c})\Big)^{-1}.\label{g8}
\end{eqnarray}
\end{widetext}

\begin{figure}
\includegraphics[width=8cm]{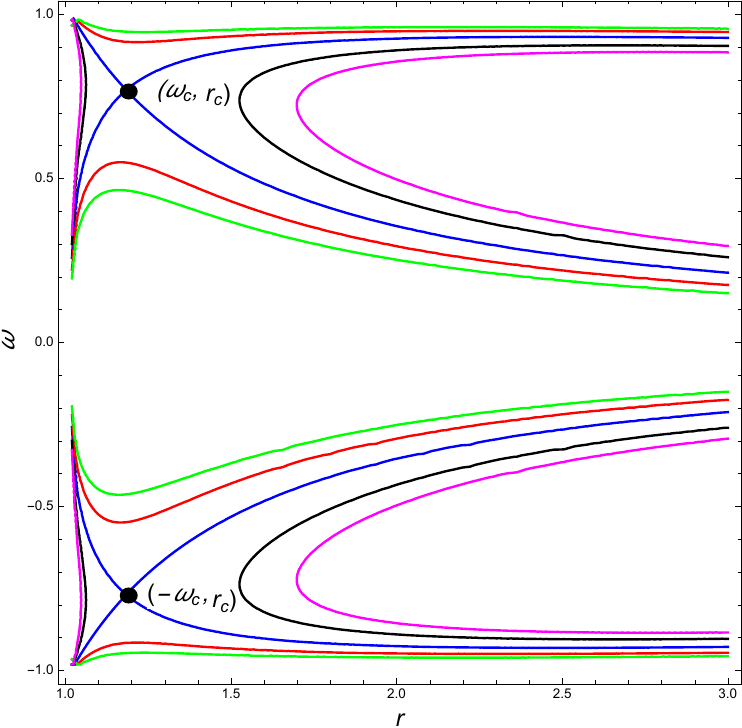}
\caption{The contour plot $\mathcal{H}$ (\ref{g5}) for the polytropic fluid with the BH parameters $m = 1$, $Q_{m} = 1$, $\nu = 0.01$, $ \Lambda _{\text{eff}}= -1$, $\alpha = 1$. The sonic (critical) points $(r_{c},\pm\omega_{c})$ are presented by black dots in a given plot. In the above-mentioned figure, five plots are given, with the colors blue, red, green, and magenta corresponding with the given Hamiltonian values $\mathcal{H}=\mathcal{H}_{c}=69.28$, $\mathcal{H}=\mathcal{H}_{c}+0.15$, $\mathcal{H}=\mathcal{H}_{c}+30$, $\mathcal{H}=\mathcal{H}_{c}-12$, and $\mathcal{H}=\mathcal{H}_{c}- 18$}.
\end{figure}

In order to prevent the Hamiltonian (\ref{g5}), from diverging, the solution curve must not cross the $r$-axis at the
point where $\omega = 0$ and $r = r_{h}$.  The dynamical system of Eqs. (\ref{g7}) and (\ref{g8}), has numerical solutions,
which are depicted in \textbf{Fig 7}. One may observe that there is only one critical point, also referred to as a saddle
point, in the accretion of a polytropic fluid. Furthermore, as shown in \textbf{Fig. 7}, the mobility of polytropic fluids
is identical to that of isothermal fluids with $k = 1/ 2$ (see Fig. 3), $k = 1/3$ (see Fig. 4), and $k = 1/4$ (see Fig. 5).
\section{Black hole mass accretion rate}
In this section, we are interested in determining the mass accretion rate of the magnetically charged EH with scalar hair BH. We are aware that when matter accretes in the vicinity of compact objects, its mass varies with the passage of time. The given relationship $\dot{M}=-\int T^{1}_{0}dS$, where $dS=\sqrt{-g}d\theta d\phi$, can be used to determine the rate of change of mass of the BH over time. To calculate the mass accretion rate, we can use the general formula, which can be found as \cite{b6}
\begin{equation}
\dot{M}=4\pi LM^{2}(e+p) ,\label{h1}
\end{equation}
By using Eqs. (\ref{a3}) and (\ref{a4}), one can get
\begin{equation}
r (\nu +r)uh \sqrt{{\cal F}(r)+u}=K_{0}.\label{h2}
\end{equation}
The relativistic flux equation can be used in the above equation to calculate
\begin{equation}
r (\nu +r)u^{r} e^{\int^{e}_{e_{\infty}}\frac{de^{'}}{e^{'}+p(e^{'})}}=-K_{1}, \label{h3}
\end{equation}
where $K_{1}$ denotes the integration constant. Inserting $p=ke$ in Eq. (\ref{h3}), we obtain
\begin{equation}
e=\Big(\frac{K_{1}}{r (\nu +r)u}\Big)^{k+1}\label{h4}.
\end{equation}
By using Eqs. (\ref{e3}), (\ref{h4}) and (\ref{h2}), we have
\begin{eqnarray}
&&u^{2}-\frac{K_{0}^{2}K_{1}^{-2(k+1)}}{(k+1)^{2}}\Big(\sqrt{r (\nu +r)}\Big)^{4k}(-u)^{2k}\nonumber\\
&&+\Big((1-\frac{2 m}{r}+\frac{Q_m^2}{r^2}-\frac{2 \alpha Q_m^4}{5 r^6}-\frac{\Lambda _{\text{eff}}r^2}{3})\nonumber\\
&&+\nu(\frac{m}{r^2}+\frac{6 \alpha Q_m^4}{5 r^7}-\frac{Q_m^2}{r^3}-\frac{\Lambda _{\text{eff}}r}{3})\Big)=0.\label{h5}
\end{eqnarray}

By solving the above equation, we can analytically determine $u^{r}$ for specific values of $k$. For ultra-stiff fluid Eq. (\ref{h5}), reduce to
\begin{equation}
u=\pm 2K_{1}\sqrt{\frac{{\cal F}(r)}{K_{0}^{2}\Big(r (\nu +r)\Big)^{2}-4K_{1}^{4}}}.\label{h6}
\end{equation}
From Eqs. (\ref{h6}),(\ref{h4}) and (\ref{h1}), we have
\begin{equation}
\dot{M}=2\pi \frac{K_{0}^{2} \Big(r (\nu +r)\Big)^{2}-4K_{1}^{4}}{\Big(r (\nu +r)\Big)^{2}{\cal F}} ,\label{h7}
\end{equation}
where $\dot{M}$ is the mass accretion rate for ultra-stiff fluid of magnetically charged EH with scalar hair as well as $K_{0}$ and $K_{1}$ are constants. Similarly, if we set $k=1/2$, we obtain the following expression
\begin{widetext}
\begin{eqnarray}
\dot{M}&=&810\pi \sqrt{5}\Big(-K _1^4\Big(r (\nu +r)(\sqrt{5} K_1^3 \Big(\Big(27 K_1^6(5 r^8 \Lambda _{\text{eff}} (\nu +r)+15 r^4 Q_m^2 (\nu -r)+6 \alpha  Q_m^4 (r-3 \nu )\nonumber\\
&&-15 r^5(m \nu -2 m r+r^2))+20 K_0^4 r^9 (\nu +r)^2\Big)\Big(K_1^6 r^7\Big)^{-1}\Big)^{1/2}+10 K_0^2 r (\nu +r))\Big)^{-1}\Big)^{3/2}.\label{h8}
\end{eqnarray}
The graph of the mass accretion rate is depicted in \textbf{Fig. 8}. From \textbf{Fig. 8}, it can be seen that as the values of the BH parameters increase, the mass accretion rate decreases.
\end{widetext}

\begin{figure*}
\subfigure[\label{fig:a}]{\includegraphics[width=8cm]{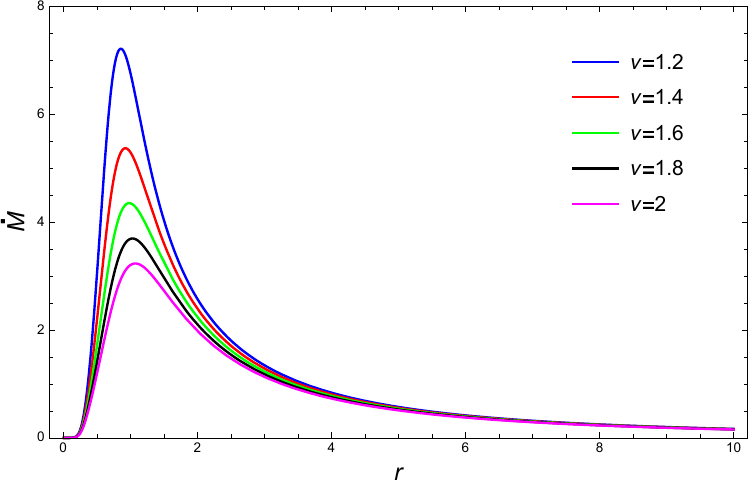}}
\subfigure[\label{fig:b}]
{\includegraphics[width=8cm]{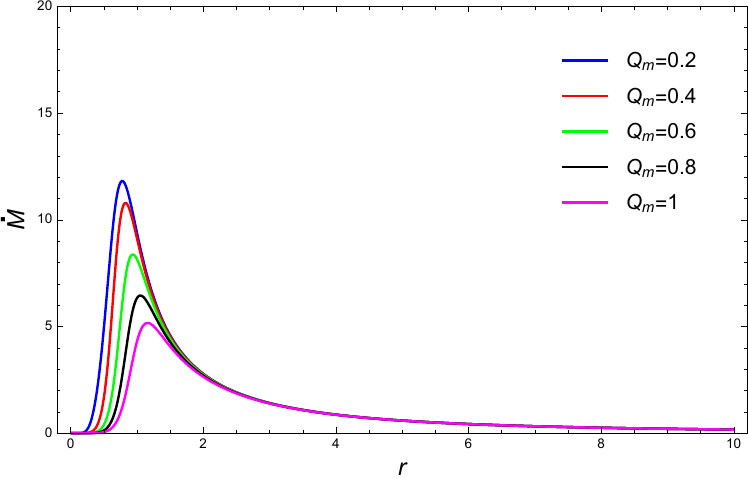}}
\subfigure[\label{fig:c}]
{\includegraphics[width=8cm]{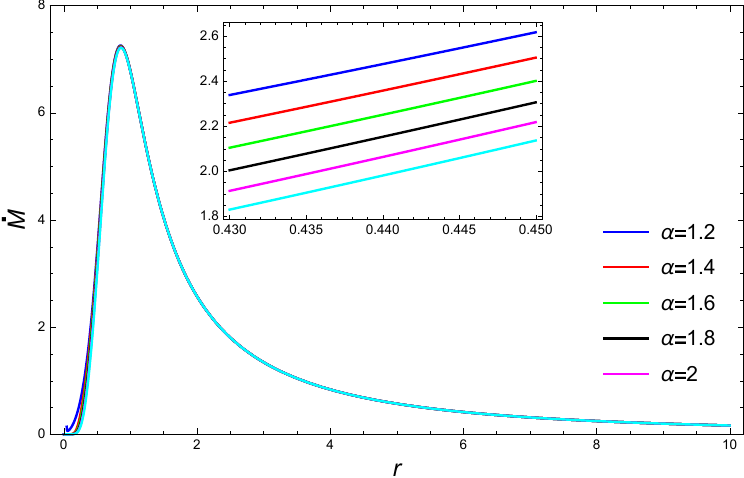}}
\caption{The graph is plotted between $\dot{M}$ verses $r$ (a) for $m=1$, $Q_{m}=0.2$, $k=1$, $K_{0}=0.1$, $\Lambda _{\text{eff}}=-1$ and various values of $\nu$ (b) $m=1$, $\nu =1$, $k=1$, $K_{0}=0.1$, $\Lambda _{\text{eff}}=-1$ and altered values of $Q_{m}$ (c) $m=1$, $Q_{m}=0.2$, $K_{0}=0.1$, $k=1$, $\Lambda _{\text{eff}}=-1$, $\nu =1.2$ and different values of $\alpha$. }.
\end{figure*}

\section{Conclusion}

This paper studies the spherical accretion flow of a perfect fluid around the magnetically charged Euler-Heisenberg BH with scalar hair. We establish two basic formulas for the analysis of the accretion processes using energy and particle conservation equations. Then, utilizing these fundamental equations, we explore the various subcategories of perfect fluids, such as isothermal fluids of ultra-stiff, ultra-relativistic, radiation, sub-relativistic, and perfect fluids. \textbf{Figures} 2, 3, 4, 5,6 and 7 provide a graphical representation of the behavior of the fluids flow near the magnetically charged EH BH. {It is essential to understand that for ultra-stiff fluids, the sonic point cannot be observed. The critical values for ultra-relativistic, radiation, and sub-relativistic fluids are given in tables. We employed the equation of state to analyze the physical properties of matter, including ultra-stiff, ultra-relativistic, radiation, and sub-relativistic fluids, in order to discover the nature of BH. In addition to the equation of state and model parameters, different types of accreting fluid exhibit varying accretion characteristics, including subsonic, supersonic, and transonic. We noticed precisely how supersonic accretion, subsequently followed by
subsonic accretion, terminated inside the BH horizon. This relates to the fluid movement around the magnetically charged
Euler-Heisenberg BH with scalar hair.} The mass accretion rate of
BH has been determined for ultra-stiff fluid ($k = 1$) and ultra-relativistic fluid ($k = 1/2$). It also graphically
investigated how the BH accelerating parameter affects the spherical mass accretion rate of perfect fluid onto gravitational
bodies. From \textbf{Fig. 8}, we observe that the mass accretion rate attains its maximum value for small radii and then
decreases to its minimum for large radii. Furthermore, we can see that the mass accretion and BH parameters have an inverse
relationship. The findings presented in the previous section show that the mass accretion rate depends on the accelerating
parameter and that this influence only becomes noticeable when the parameter's value is small. We are unable to calculate
the mass accretion rate for other values of the state parameter analytically because Eq. (\ref{h5}), becomes highly non-linear
in terms of $u$, so it is not possible to find an explicit form of $\dot{M}$.

{Here we would like to mention that the study performed in this paper mainly focuses on the accretion properties of different fluids onto the Euler-Heisenberg BH with the scalar hair. It is interesting to extend the current work to other related astrophysical processes, such as the black hole images with different accretion processes or thin accretion disks, and explore how the scalar hair can affect the size and shape of the corresponding image. In particular, together with the recent observations of the black hole images by the Event Horizon Telescope and inspired by \cite{Vagnozzi:2022moj, Nozari:2023flq}, we may run a systematic study in this direction by constraining the parameters of EH BH with scalar hair in our next step. We expect to return to the above issues soon in future work.}

\section*{Acknowledgements}

The work of G. Abbas has been partially supported by the National Natural Science Foundation of China under project No. 11988101. He is grateful to the compact objects and diffused medium Research Group at NAOC led by Prof. JinLin Han for the excellent hospitality and friendly environment. He is also thankful to The Islamia University of Bahawalpur, Pakistan for the grant of 06 months the study leave. Tao Zhu is supported by the Zhejiang Provincial Natural Science Foundation of China under Grant No. LR21A050001 and LY20A050002,  the National Key Research and Development Program of China Grant No.2020YFC2201503, and the National Natural Science Foundation of China under Grant No. 12275238 and No. 11675143.

\end{document}